\documentclass[aps, superscriptaddress,reprint,amsmath,amssymb,prl]{revtex4-2}
\usepackage[]{graphicx}
\usepackage{multirow}
\usepackage{dcolumn}
\usepackage[usenames]{color}
\usepackage{booktabs}
\usepackage[version=4]{mhchem}
\usepackage[utf8]{inputenc}
\usepackage{subcaption}
\usepackage{makecell}
\usepackage{multirow}
\usepackage{array}
\usepackage{bm}
\usepackage[colorlinks=true, allcolors=blue]{hyperref}

\usepackage{caption}
\usepackage{xcolor}
\definecolor{darkgreen}{RGB}{0,150,0}

\usepackage{siunitx}
\DeclareSIUnit\angstrom{\text{Å}}
\DeclareSIUnit{\debye}{D}
\DeclareSIUnit{\calorie}{cal}
\DeclareSIUnit{\kcal}{\kilo\calorie}
\DeclareSIUnit{\step}{step}

\usepackage{booktabs} 

\begin{document}

\title{MBD-ML: Many-body dispersion from machine learning for molecules and materials}

\author{Evgeny Moerman}
\affiliation{Department of Physics and Materials Science, University of Luxembourg, L-1511 Luxembourg City, Luxembourg}

\author{Adil Kabylda}
\affiliation{Department of Physics and Materials Science, University of Luxembourg, L-1511 Luxembourg City, Luxembourg}

\author{Almaz Khabibrakhmanov}
\affiliation{Department of Physics and Materials Science, University of Luxembourg, L-1511 Luxembourg City, Luxembourg}

\author{Alexandre Tkatchenko}
\email{alexandre.tkatchenko@uni.lu}
\affiliation{Department of Physics and Materials Science, University of Luxembourg, L-1511 Luxembourg City, Luxembourg}

\begin{abstract}
\noindent 
Van der Waals (vdW) interactions are essential for describing molecules and materials, from drug design and catalysis to battery applications. These omnipresent interactions must also be accurately included in machine-learned force fields. The many-body dispersion (MBD) method stands out as one of the most accurate and transferable approaches to capture vdW interactions, requiring only atomic $C_6$ coefficients and polarizabilities as input. We present MBD-ML, a pretrained message passing neural network that predicts these atomic properties directly from atomic structures. Through seamless integration with libMBD, our method enables the immediate calculation of MBD-inclusive total energies, forces, and stress tensors. By eliminating the need for intermediate electronic structure calculations, MBD-ML offers a practical and streamlined tool that simplifies the incorporation of state-of-the-art vdW interactions into any electronic structure code, as well as empirical and machine-learned force fields.
\end{abstract}

\maketitle

\section{Introduction}

Van der Waals (vdW) dispersion interactions play a decisive role in the properties of molecular crystals, condensed matter, and biological systems~\cite{Hermann2017,Stohr2019-Review}. They govern polymorphism in organic crystals~\cite{reilly2014role, hoja2019reliable}, protein folding~\cite{stohr2019quantum}, and the cohesion of layered, polymeric, and porous materials~\cite{bjorkman2012, tawfik2018}. In many such systems, dispersion interactions are not a small perturbation but a leading contribution that critically determines relative energies, equilibrium geometries, and atomistic dynamics.

Density functional theory (DFT) has become a cornerstone of atomistic simulations of molecules and solids, and increasingly serves as a reference~\cite{barroso2024open, levine2025open, maurer2019advances} for the construction of machine-learning force fields (MLFFs)~\cite{unke2021machine}. However, most widely used semi-local and hybrid exchange-correlation functionals, as well as local ML architectures, do not account for long-range vdW dispersion. As a consequence, explicit dispersion interactions are indispensable for achieving quantitative accuracy, unless one deals with systems where strong covalent bonds dominate the desired behavior~\cite{unke2021machine, kabylda2025molecular}.

Over the past two decades, a wide range of vdW methods have been proposed. Early and still widely adopted approaches include the pairwise-additive DFT-D methods~\cite{Grimme2016}, most notably the D3 correction by Grimme~\cite{grimme2010consistent}, based on tabulated $C_6$ coefficients and polarizabilities $\alpha_0$. The Tkatchenko-Scheffler (TS)~\cite{Tkatchenko2009} method introduced an important conceptual advance by making these quantities environment-dependent through Hirshfeld partitioning of the electron density. Furthermore, in contrast to empirical methods, the TS method, as well as its further extensions, is a self-consistent interatomic vdW density functional~\cite{ferri2015electronic}. Similarly, the exchange-hole dipole moment (XDM) method~\cite{becke2007exchange, Johnson2017} derives atomic polarizabilities from the electronic density and incorporates higher-order multipole contributions to the dispersion energy through $C_8$ and $C_{10}$ coefficients obtained from the exchange-hole multipole moments.

A fundamental limitation of all pairwise schemes is their inability to capture collective many-body effects in dispersion interactions~\cite{Hermann2017,Stohr2019-Review}. This shortcoming is addressed by the many-body dispersion (MBD) method~\cite{tkatchenko2012accurate}, which models dispersion interactions using a system of dipole-coupled quantum harmonic oscillators parametrized to reproduce atomic $\alpha_0$ and $C_6$ coefficients. 
The MBD framework naturally accounts for non-additive many-body effects as well as polarization anisotropy~\cite{tkatchenko2013}, and has been shown to substantially improve accuracy for molecular crystals, layered materials, and large molecular systems~\cite{Ambrosetti2014, DiStasio2014, Hermann2017-NatComms, kim2020umbd, hermann2020density}. Several variants of MBD, differing in their parametrization and treatment of short-range screening, have been developed~\cite{Ambrosetti2014, gould2016fractionally, kim2020umbd, hermann2020density, Khabibrakhmanov2025-JACS}, including the state-of-the-art MBD-NL formulation~\cite{hermann2020density}.

Despite this rich variety of vdW methods and the demonstrated accuracy of many-body approaches, the D3 correction remains the most widely used in large-scale and high-throughput applications. This includes generation of DFT-based datasets for molecules~\cite{levine2025open} and materials~\cite{barroso2024open}, as well as routine applications in computational chemistry and materials science. While the limitations of D3 are well documented~\cite{paul2025dft, sedlak2017empirical, proppe2019gaussian, caldeweyher2019generally, witte2017assessing, tkachenko2024smoother, kim2020umbd, briccolani2026weak}, its popularity is largely driven by its ease of use. D3 is implemented as a standalone library and requires only atomic positions and element types, without the need for electronic-structure information or tight coupling to DFT codes. For the MBD method, however, the reliance on electronic-structure calculations to parametrize the MBD Hamiltonian remains a major bottleneck that limits its applicability in large-scale simulations and ML architectures. 

\begin{figure*}
    \includegraphics[width=0.9\linewidth]{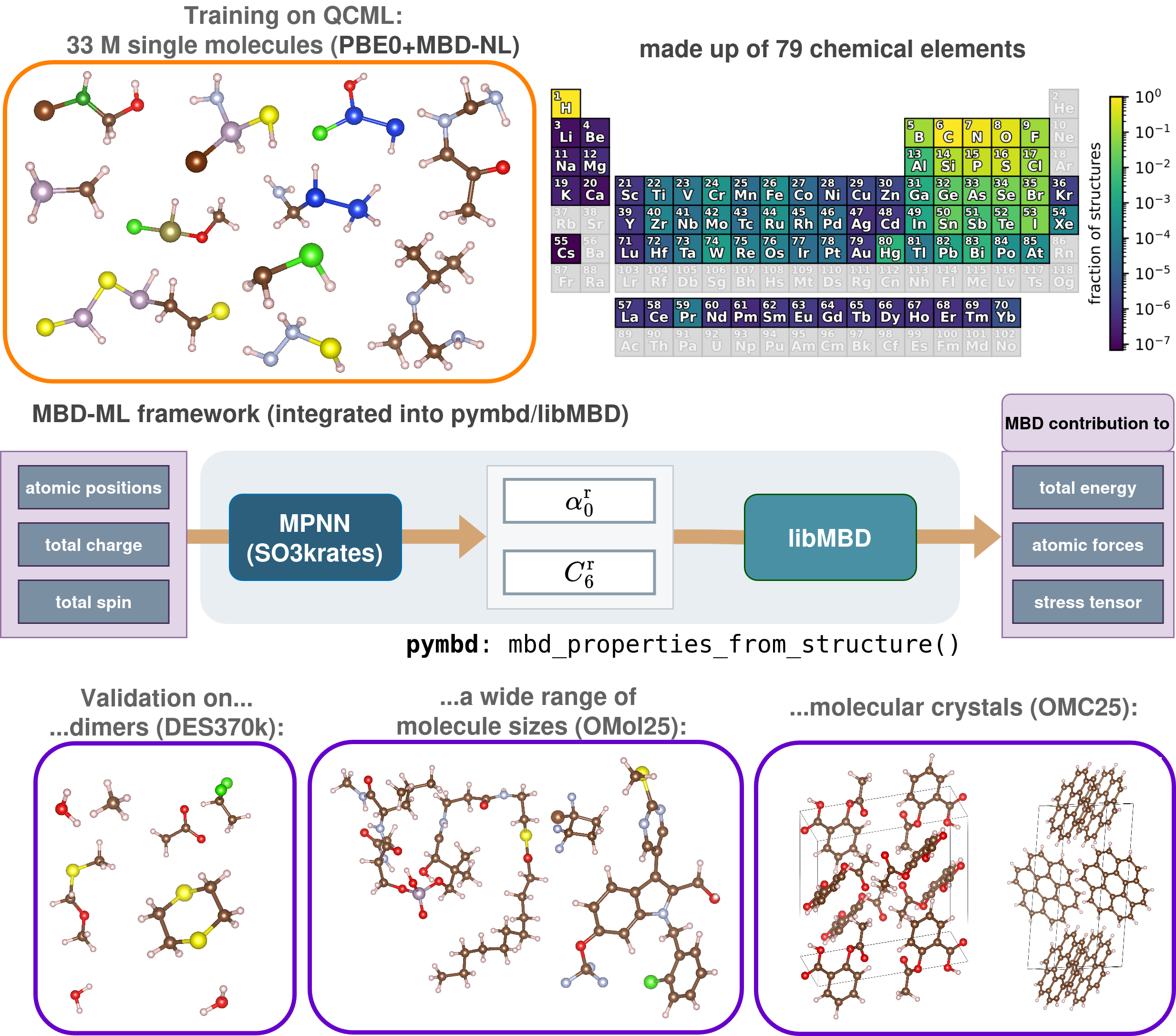}
    \caption{Overview of the training and the validation of the MBD-ML model: (top left) MBD-ML was trained on the QCML molecular data set spanning almost the entire periodic system (top right; Reproduced from Ref.~\citenum{ganscha2025qcml} under CC BY 4.0). (middle) The model has been integrated into the libMBD python interface, pymbd, to allow direct calculation of MBD properties from the atomic structure. (bottom) The  model was validated on both molecules and molecular crystals in predicting fundamental properties including total energy and  force contributions and -- in the case of organic crystals -- to perform geometry optimizations and recover the energy ranking of different polymorphs}
    \label{fig:overview}
\end{figure*}

In this work, we remove this limitation for the MBD method by introducing a machine learning-based framework, termed MBD-ML, that enables accurate many-body dispersion calculations using only atomic geometry as input. We achieve this by training a state-of-the-art message-passing neural network to predict the effective atomic polarizabilities $\alpha_0$ and $C_6$ coefficients, using the DFT+MBD-NL method~\cite{hermann2020density} as a reference. By integrating this model into the libMBD library~\cite{hermann2023libmbd} (see Fig.~\ref{fig:overview}), we make it possible to compute dispersion energies, atomic forces, and stress tensors at the MBD-NL level of accuracy without performing any electronic-structure calculations. Previous machine-learned MBD models~\cite{poier2022accurate} employed simple deep-learning architectures to predict Hirshfeld volume ratios for the antecedent MBD@rsSCS method and were limited to small molecules containing C, H, N, and O. In contrast, the present MBD-ML model is applicable to both molecules and crystals across over 70 chemical elements and is directly integrated into the libMBD infrastructure. Moreover, $\alpha_0$ and $C_6$ ratios from MBD-NL, which we used for training, are more robust and transferable across chemical space, facilitating treatment of ionic and metallic compounds, which are problematic for MBD@rsSCS relying on Hirshfeld volume ratios~\cite{gould2016fractionally, bryenton2023many}.

We demonstrate the transferability and general performance of the MBD-ML framework on subsets of five state-of-the-art data sets (see Fig.~\ref{fig:overview}) covering small molecules of widely varying compositions (QCML~\cite{ganscha2025qcml}) and sizes (OMol25~\cite{levine2025open}), biomolecular and  drug-like dimers (DES370k~\cite{donchev2021quantum}), molecular crystals (OMC25~\cite{gharakhanyan2025open}) and inorganic materials (OMat24~\cite{barroso2024open}). Beyond that we assess the robustness and practical accuracy of MBD-ML by applying the method to the prediction of structures and relative energetics of molecular-crystal polymorphs, and comparing the results directly against reference \emph{ab initio} MBD-NL calculations. Our results demonstrate that MBD-ML retains the accuracy of MBD-NL, while dramatically extending its usability to large-scale ML-based data generation or atomistic modeling. 

\section{Methods}
\subsection{The MBD framework}\label{sec:mbd-nl-method}

The MBD method~\cite{tkatchenko2012accurate} employs coupled quantum Drude oscillators (QDOs)~\cite{Jones2013, Khabibrakhmanov2025-JCP} to model collective electron density fluctuations, underpinning vdW dispersion. In this framework, the electronic response of each atom is represented by a single QDO, which captures the collective fluctuations of all valence electrons. Each QDO consists of a negatively charged quasiparticle harmonically bound to a positively charged pseudonucleus. The mapping between atoms and oscillators is established via the static polarizability $\alpha_{0,i}$ and effective excitation frequency $\omega_i = 4C_{6,ii}/3\alpha_{0,i}^2$. 

The interaction between oscillators is described via the long-range dipole-dipole coupling tensor $\mathbf{T}^{\mathrm{lr}}$. The resulting Hamiltonian for a system of $N$ coupled QDOs is given by
\begin{align}\label{eq:mbd-h}
    H_{\mathrm{MBD}}(\{&\mathbf{R}_i,\alpha_{0,i},\omega_i\})
    =
    \sum_{i}-\frac{1}{2}\nabla_{\boldsymbol{\xi}_i}^2 + \nonumber
    \frac{1}{2}\omega_i^2 \boldsymbol{\xi}_i^2 \\
    &+\frac{1}{2}\sum_{ij}\omega_i\omega_j\sqrt{\alpha_{0,i}\alpha_{0,j}}
    \boldsymbol{\xi}_i\mathbf{T}^{\mathrm{lr}}_{ij}\boldsymbol{\xi}_j \ ,
\end{align}
with $\boldsymbol{\xi}_i = \sqrt{m_i}\mathbf{x}_i$ denoting the mass-weighted displacements of the oscillating charge relative to the nuclei position $\mathbf{R}_i$. 

Since the Hamiltonian~\eqref{eq:mbd-h} is quadratic in displacements, the many-body dispersion energy can be obtained by direct diagonalization:
\begin{equation}
    E_{\mathrm{MBD}} = 
    \sum_{k=1}^{3N} \frac{\tilde{\omega}_k}{2} - \sum_{i=1}^N \frac{3\omega_i}{2}
    \mathrm{,}
\end{equation}
where $\tilde{\omega}_k$ are the frequencies of the $3N$ collective MBD eigenmodes. The resulting MBD energy is formally equivalent to the adiabatic-connection fluctuation–dissipation (ACFD-RPA) correlation energy~\cite{tkatchenko2013}, while the Hamiltonian diagonalization provides a numerically efficient route to its evaluation. The computational complexity of the MBD method scales as $\mathcal{O}(N^3)$, matching that of Kohn-Sham DFT. However, we emphasize that the prefactor is substantially smaller, such that the cost of an MBD calculation constitutes only a small fraction of a single DFT self-consistency cycle, even for large systems~\cite{hermann2023libmbd}.

Within this general formalism, several variants of the MBD method have been developed, which primarily differ in the determination of the atomic polarizabilities $\alpha_{0,i}$, dispersion coefficients $C_{6,ii}$, and the long-range interaction tensor $\mathbf{T}^{\mathrm{lr}}$. Among these, the non-local MBD approach (MBD-NL)~\cite{hermann2020density} represents the most advanced and robust formulation, exhibiting uniformly high accuracy across covalent, ionic, and metallic systems. This is achieved by using the Vydrov-Van Voorhis (VV) polarizability functional~\cite{vydrov2010dispersion} to obtain $\alpha_{0,i}^{\mathrm{VV}}$ and $C_{6,ii}^{\mathrm{VV}}$ coefficients that parametrize the MBD Hamiltonian. 

A key feature of MBD-NL is the explicit removal of contributions from jellium-like density regions~\cite{hermann2020density}. This cutoff is essential to avoid double counting of correlation energy, as most exchange-correlation functionals are exact for the homogeneous electron gas by construction. Furthermore, to compensate for the approximate nature of the VV polarizability functional, the resulting $\alpha_{0,i}^{\mathrm{VV}}$ and $C_{6,ii}^{\mathrm{VV}}$ are renormalized using high-quality free-atom reference data:
\begin{equation}
    \alpha_{0,i} = \frac{\alpha_{0,i}^{\mathrm{VV}}}{\alpha_{0,i}^{\mathrm{VV,free}}} \alpha_{0,i}^{\mathrm{ref,free}} \ ,
    \quad
    C_{6,ii} = \frac{C_{6,ii}^{\mathrm{VV}}}{C_{6,ii}^{\mathrm{VV,free}}} C_{6,ii}^{\mathrm{ref,free}} \ ,
\end{equation}
where $\alpha_{0,i}^{\mathrm{ref,free}}$ and $C_{6,ii}^{\mathrm{ref,free}}$ denote the corresponding free-atom reference values. Further technical details of the MBD-NL formalism and its implementation can be found in the original MBD-NL publication~\cite{hermann2020density}.

The combination of the VV polarizability functional, the jellium cutoff, and normalization to free-atom reference data yields a robust and transferable parametrization of the MBD Hamiltonian across diverse chemical environments. In particular, the MBD-NL formulation avoids the ``polarization catastrophe'' observed for earlier MBD variants in transition-metal compounds~\cite{gould2016fractionally, bryenton2023many} and significantly reduces overbinding error in metals and ionic solids~\cite{hermann2020density}, achieving balanced accuracy across the periodic table.

\subsection{The MBD-ML framework}
Building on the MBD-NL formalism described above, we now introduce the MBD-ML model, which learns to reproduce its key atomic inputs, $\alpha_0^{\mathrm{r}}$ and $C_6^{\mathrm{r}}$, directly from the atomic configuration.
MBD-ML utilizes the SO3krates architecture~\cite{frank2024euclidean}, an equivariant message passing neural network that represents a general and flexible framework for training state-of-the-art MLFFs. A closely related model is the recent SO3LR~\cite{kabylda2025molecular}, which was trained to map atomic positions $R$, atomic numbers $Z$ and the total charge $Q$ and spin $S$ to atomic forces $\bm{F}_i$, partial charges $q_i$ and Hirshfeld ratios $h_i$. Here, we modified and re-trained the model to predict the $\alpha_0^{\mathrm{r}}$ and $C_6^{\mathrm{r}}$ ratios:
\begin{equation}
     \alpha_0^{\mathrm{r}}, C_6^{\mathrm{r}} = \mathrm{SO3krates}(R, Z, Q, S) \ \text{.}
\end{equation}
defined as the ratios between the $C_{6,ii}^{\mathrm{VV}}$ and $\alpha_{0,i}^{\mathrm{VV}}$ atomic coefficients computed by the VV polarizability functional in a multi-atomic system and their free-atom equivalents $C_{6,ii}^{\mathrm{VV,free}}$ and $\alpha_{0,i}^{\mathrm{VV,free}}$,
\begin{equation} \label{eq:def_ratios}
    \alpha_{0,i}^{\mathrm{r}} = \frac{\alpha_{0,i}^{\mathrm{VV}}}{\alpha_{0,i}^{\mathrm{VV,free}}}
    \mathrm{,}
    \quad
    C_{6,ii}^{\mathrm{r}} = \frac{C_{6,ii}^{\mathrm{VV}}}{C_{6,ii}^{\mathrm{VV,free}}} \ 
    \mathrm{.}
\end{equation}
which in the following will be simply referred to as $C_6^{\mathrm{r}}$ and $\alpha_0^{\mathrm{r}}$ for brevity.

In contrast to polarizabilities and dispersion coefficients, broadly varying in magnitude across the periodic table, these unit-less ratios are typically confined in the range $\approx 0-2$ for most elements and chemical environments, which makes $\alpha_0^{\mathrm{r}}$ and $C_6^{\mathrm{r}}$ naturally suitable targets for machine learning. We note that the MBD-ML model predicts $\alpha_0^{\mathrm{r}}$ and $C_6^{\mathrm{r}}$ separately, while within the SO3LR model the analogous quantities are directly connected ($\alpha_{0,i}^{\mathrm{r}} = h_i$ and $C_{6,i}^{\mathrm{r}} = h_i^2$). This important distinction originates from the underlying MBD-NL method and makes the resulting MBD predictions more robust for general systems, as it was explained above.

\begin{table*}
\caption{Performance of MBD-ML models on ratios and MBD properties for different test sets in terms of RMSE and MAE values. Errors for $\alpha_0^{\mathrm{r}}$ and $C_6^{\mathrm{r}}$ are unitless. $N$ denotes the number of systems in each test set. \textsuperscript{a}Excluding negatively charged molecules (due to insufficient accuracy of \emph{ab initio} treatment; see QCML discussion in the \emph{Model validation} section). \textsuperscript{b}Excluding alkali and alkaline earth metals (due to under-representation in the training set; see DES370k discussion in the \emph{Model validation} section)}
\centering
\setlength{\tabcolsep}{5pt}
\begin{tabular}{l cccccccccc}
\toprule
& \multicolumn{2}{c}{\makecell{QCML\textsuperscript{a}\\($N{=}83226$)}} 
& \multicolumn{2}{c}{\makecell{DES370k\textsuperscript{a,b}\\($N{=}309130$)}} 
& \multicolumn{2}{c}{\makecell{OMOL25\textsuperscript{b}\\($N{=}825$)}} 
& \multicolumn{2}{c}{\makecell{OMC25\\($N{=}200$)}} 
& \multicolumn{2}{c}{\makecell{OMAT24\\($N{=}203$)}}\\
\\[-8pt]
& MAE & RMSE & MAE & RMSE & MAE & RMSE & MAE & RMSE & MAE & RMSE\\

\midrule
\multicolumn{11}{l}{\textit{Ratio Predictions}}\\[2pt]
\quad $\alpha_0^{\mathrm{r}}$ & 0.013 & 0.020 & 0.014 & 0.025 & 0.023 & 0.033 & 0.025 & 0.031 & 0.247 & 0.429\\
\quad $C_6^{\mathrm{r}}$      & 0.014 & 0.023 & 0.013 & 0.022 & 0.021 & 0.030 & 0.017 & 0.022 & 0.269 & 0.895\\
\midrule
\multicolumn{11}{l}{\textit{MBD Properties}}\\[2pt]
\quad $E_{\mathrm{MBD}}$ (meV/atom)   & 0.158 & 0.228 & 0.117 & 0.165 & 0.198 & 0.226 & 0.713 & 0.841 & -- & --\\
\quad $F_{\mathrm{MBD}}$  (meV/\AA)  & 0.302 & 0.440 & 0.266 & 0.387 & 0.564 & 0.707 & 0.786 & 0.930 & -- & --\\
\quad $\sigma_{\mathrm{MBD}}$ (meV/\AA$^3$) & --    & --    & --    & --    & --    & --    & 0.112 & 0.143 & -- & --\\
\bottomrule
\end{tabular}
\label{tab:performance-cs-models}
\end{table*}

The values of $\alpha_{0,i}$ and $C_{6,ii}$, required to parametrize the MBD Hamiltonian~\eqref{eq:mbd-h}, are obtained by multiplying the predicted ratios by the free-atom reference quantities according to Eq.~\eqref{eq:def_ratios}. This is done internally within the libMBD interface and hence does not require any user intervention. Thus, the MBD-ML-predicted ratios are sufficient to directly compute all MBD-related quantities, including energies, atomic forces, and stress tensors, via the efficient libMBD routines~\cite{hermann2023libmbd}. 

Training on the atomic ratios rather than directly on the MBD properties also offers another key advantage -- $\alpha_0^{r}$ and $C_6^{r}$ ratios are relatively short-ranged quantities, whereas energies and forces can be very sensitive to long-range interactions. Consequently, a semi-local message-passing architecture like SO3krates is well-suited for this application. In addition, the $\alpha_0^{\mathrm{r}}$ and $C_6^{\mathrm{r}}$ ratios are not very sensitive to the underlying density functional approximation (DFA), so that application of the model in combination with other exchange-correlation functionals only requires the modification of the damping parameter $\beta$, governing the short-range behavior of the dipolar tensor $\mathbf{T}^{\mathrm{lr}}$. In the present work, $\beta$ was not re-optimized for the MBD-ML model and the tabulated default values were used for the respective DFAs ($\beta_{\mathrm{PBE}} = 0.81,\, \beta_{\mathrm{PBE0}} = 0.83$) following Ref.~\cite{hermann2020density}.  

The current MBD-ML implementation uses SO3krates with two message-passing layers and a cutoff radius of $4\,\mathrm{\AA}$, giving an effective receptive field of $8\,\mathrm{\AA}$, and was trained on the QCML dataset~\cite{ganscha2025qcml}, comprising over 30 million molecules with up to 8 non-hydrogen atoms across 79 chemical elements.
QCML provides structures, energies, and forces at the PBE0+MBD-NL level, along with MBD-NL polarizabilities $\alpha_{0,i}^{\mathrm{VV}}$, and $C_{6,ii}^{\mathrm{VV}}$ coefficients, enabling direct training of the MBD-ML model. We refer the reader to Section~S1 of the Supplementary Information for the technical details of the training procedure.
    
\section{Model validation}\label{sec:model-validation}
To quantify the accuracy and identify potential limitations of the MBD-ML model in predicting MBD corrections to total energies, atomic forces, and stress tensors (for periodic systems), we first evaluate the trained model on three test sets spanning different chemical compound spaces. Beyond the hold-out validation on the QCML data, we assess performance on molecular non-covalent dimers from DES370k~\cite{donchev2021quantum}, and on molecular crystals using a subset of OMC25~\cite{gharakhanyan2025open}. Details on the selection and recalculation of the OMol25 and DES370k subsets can be found in Section~S2 and S3, respectively.

\subsection{Molecular data sets}

\begin{figure*}
\centering
\begin{subfigure}[t]{0.245\textwidth}
\caption{}
\includegraphics[width=\linewidth]{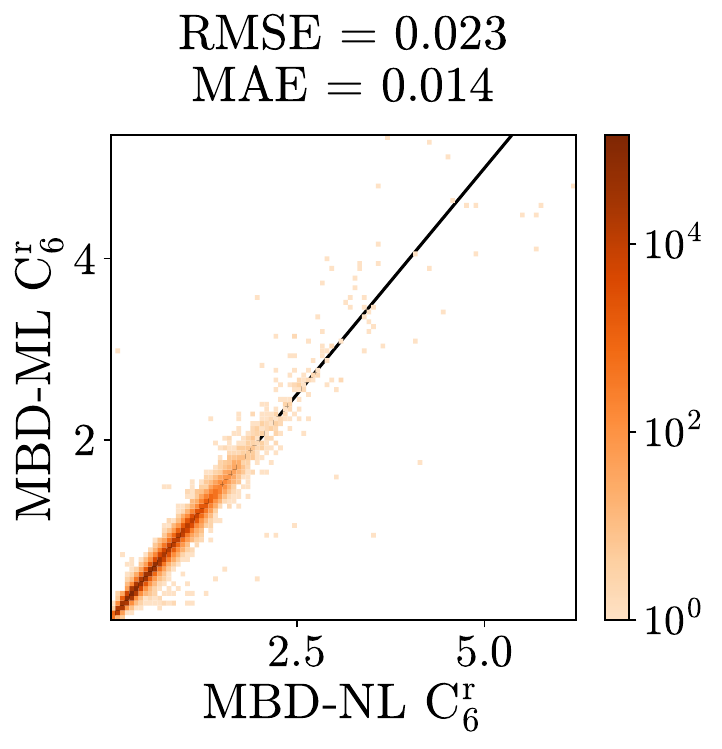}
\end{subfigure}%
\begin{subfigure}[t]{0.245\textwidth}
\caption{}
\includegraphics[width=\linewidth]{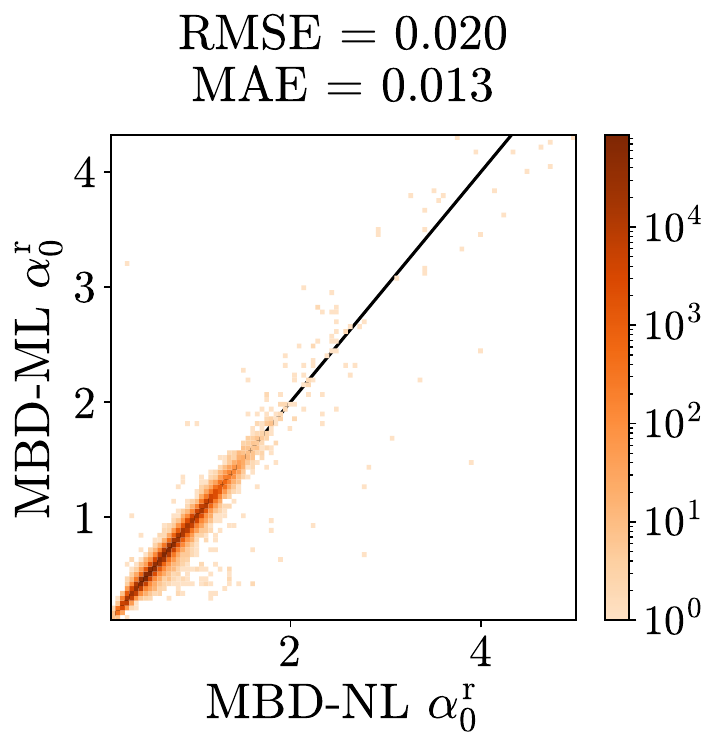}
\end{subfigure}%
\begin{subfigure}[t]{0.245\textwidth}
\caption{}
\includegraphics[width=\linewidth]{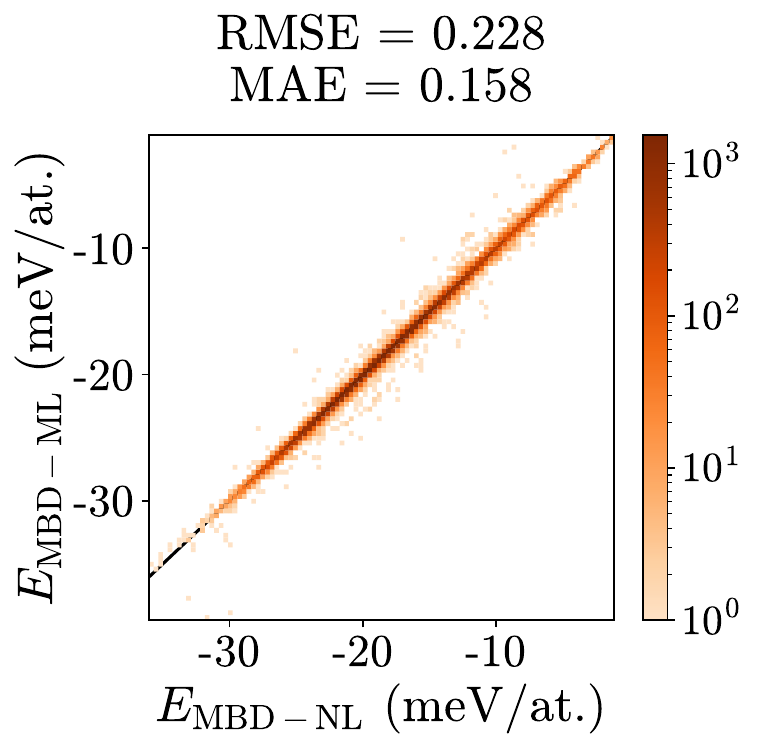}
\end{subfigure}%
\begin{subfigure}[t]{0.245\textwidth}
\caption{}
\includegraphics[width=\linewidth]{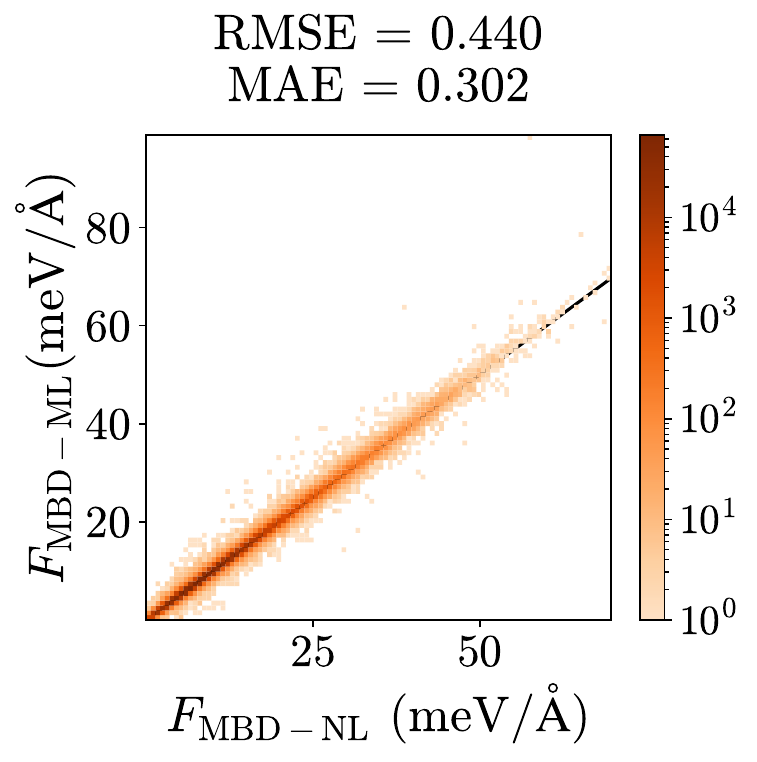}
\end{subfigure}
\caption{Performance of PBE0+MBD-ML in predicting the $C_6$
and $\alpha_0$ ratios and the MBD contribution to the total energy and atomic forces of the QCML test set}
\label{fig:c6-a0-e-f-performance-qcml}
\end{figure*}

\textbf{QCML.} The MBD-ML model trained on the QCML dataset achieves strong performance for neutral and positively charged molecules (Table~\ref{tab:performance-cs-models}, Fig.~\ref{fig:c6-a0-e-f-performance-qcml}), with RMSEs of 0.023 and 0.020 for $\alpha_0^{\mathrm{r}}$ and $C_6^{\mathrm{r}}$ ratios, respectively. These low errors demonstrate that MBD-ML accurately captures the quantum mechanical effects that modulate atomic polarizabilities and dispersion coefficients in diverse chemical environments. The accurately predicted ratios yield sub-meV errors in the MBD energy ($0.229\,\text{meV/atom}$) and force ($0.440\,\text{meV/atom}$) contributions (Fig.~\ref{fig:c6-a0-e-f-performance-qcml}), indicating its suitability as an accurate drop-in replacement of the \emph{ab initio} MBD-NL method for routine electronic-structure or MLFF calculations.

Additionally, validation on the hold-out test set revealed that some negatively charged molecules are inadequately described by the employed \emph{ab initio} treatment. The $\alpha_0^{\mathrm{r}}$ and $C_6^{\mathrm{r}}$ ratios vary substantially with increasing confinement potential cutoff radius in \textsc{FHI-aims} (Section~S4.1), indicating that the underlying VV polarizability functional is highly sensitive to electronic density tails. This sensitivity arises because additional electrons in anions are often weakly bound and extremely delocalized, resulting in a far-reaching, slowly decaying electron density. Moreover, our random inspection of several pathological cases revealed that all of those anions had negative electron affinities, meaning that they are not stable thermodynamically (please consult Section~S4 for the results). For such systems, a bound solution only exists because of the finite basis set span. For these reasons, we decided to exclude negatively charged molecules from the validation, and we do not recommend using the QCML data for their vdW energies and ratios until rigorous \textit{ab initio} benchmarks and stability-based filtering are performed. Interestingly, including anionic molecules deteriorates prediction accuracy for MBD ratios (Section~S4.2), with errors exceeding one order of magnitude, yet MBD energies and forces remain nearly unchanged (Fig.~S5).

Beyond these well-known issues of converging the electronic structure of molecular anions, charged systems are generally challenging for the state-of-the-art vdW methods (MBD-NL, XDM, DFT-D4)~\cite{nickerson2023comparison}, in particular, charged-neutral interactions. For the MBD case, the observed interaction energy disagreements actually arise from the underlying DFA and can be mitigated e.g. by combining MBD with density-corrected DFAs~\cite{zhao2025accurate}.

\begin{figure*}
\centering
\begin{subfigure}{0.3\linewidth}
\caption{}
\includegraphics[width=\linewidth]{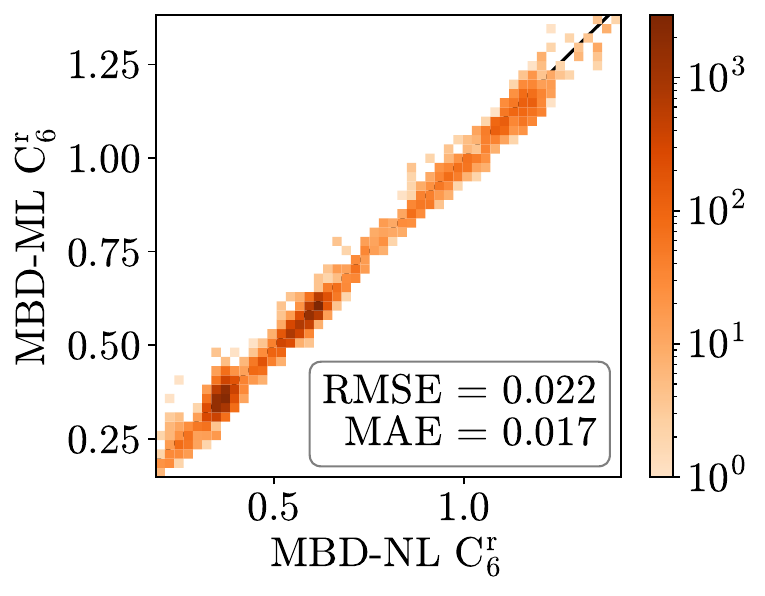}
\end{subfigure}%
\begin{subfigure}{0.3\linewidth}
\caption{}
\includegraphics[width=\linewidth]{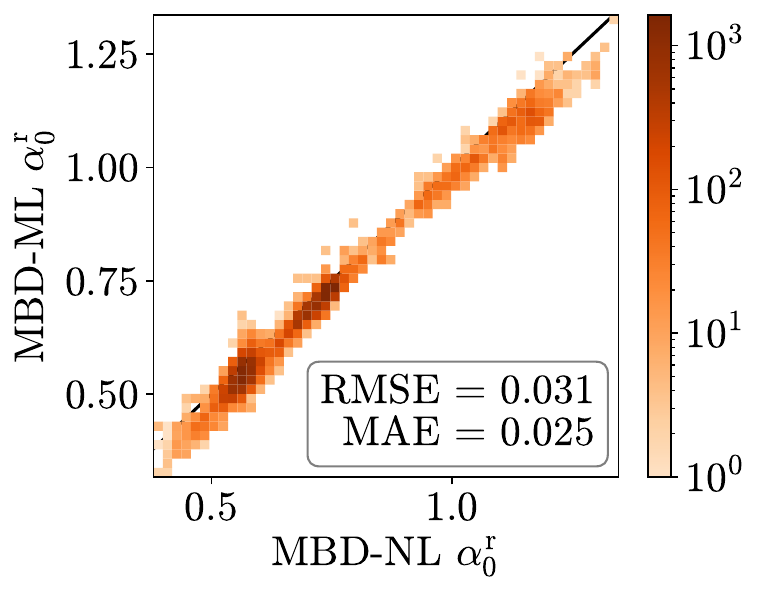}
\end{subfigure}%

\begin{subfigure}{0.3\linewidth}
\caption{}
\includegraphics[width=\linewidth]{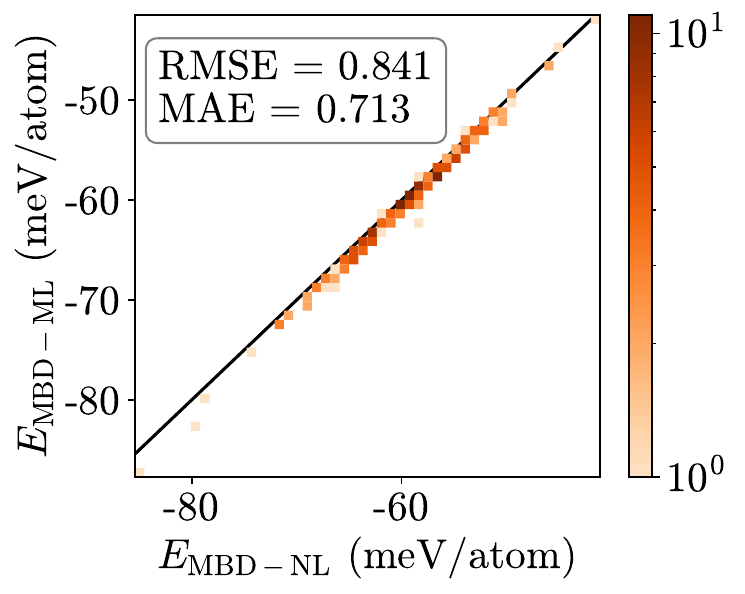}
\end{subfigure}%
\begin{subfigure}{0.3\linewidth}
\caption{}
\includegraphics[width=\linewidth]{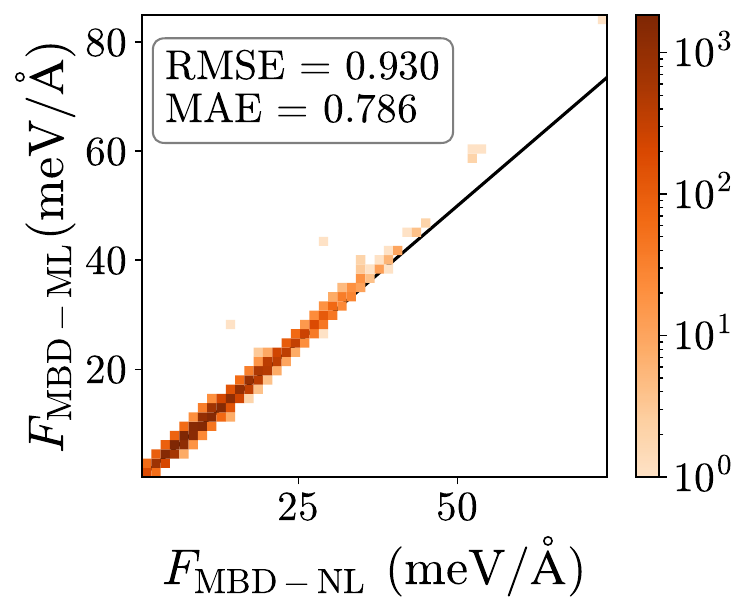}
\end{subfigure}%
\begin{subfigure}{0.3\linewidth}
\caption{}
\includegraphics[width=\linewidth]{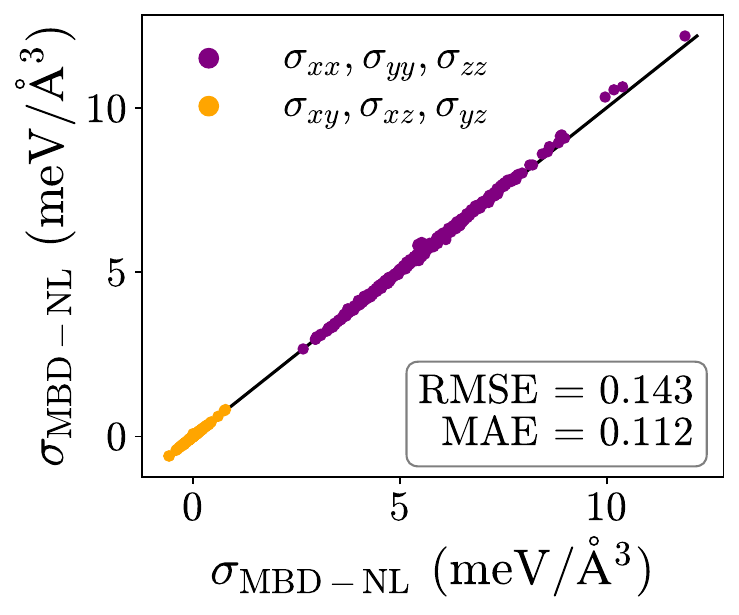}
\end{subfigure}
\caption{Performance of PBE0+MBD-ML in predicting the $C_6$ and $\alpha_0$ ratios and the MBD contribution to the total energy, atomic forces and stress tensor components in the OMC25 test set}
\label{fig:c6-a0-e-f-s-performance-omc25}
\end{figure*}

\textbf{DES370k.} To validate the MBD-ML model beyond single molecules, we tested it on molecular dimers from the DES370k dataset~\cite{donchev2021quantum}, recomputed at the PBE0+MBD-NL level as a part of the QCell biomolecular dataset~\cite{kabylda2025qcell}. For 309,130 dimers excluding negatively charged systems and structures containing alkali or alkaline earth elements (see below), MBD-ML achieves prediction errors for MBD ratios comparable to the QCML hold-out set, with slightly improved accuracy for MBD energies and forces (RMSEs of $0.165\,\text{meV/atom}$ and $0.387\,\text{meV/\AA}$, respectively; see Table~\ref{tab:performance-cs-models}). Given the substantial differences in chemical compound space between QCML (diverse small molecules) and DES370k (biomolecular and drug-like dimers), the MBD-ML model proves transferable and reliable beyond its training distribution.

A more in-depth analysis of the validation results shows that accuracy substantially deteriorates for structures containing alkali (Li, Na, K) and alkaline earth elements (Be, Mg, Ca), which are severely under-represented in the QCML training set~\cite{ganscha2025qcml}. Each element appears in only $10^{-5} - 10^{-4}\%$ of the 33.5 million QCML structures (approximately $3 - 30$ structures per element), making them the rarest elements alongside the lanthanides. Consequently, we exclude these elements from our assessment. A complete overview of MBD-ML performance including these problematic cases is provided in Section~S5. Hence, we recommend to use the default version of MBD-ML with care when dealing with such under-represented elements. However, this limitation can be overcome by fine-tuning the model on a specific dataset of interest. 

\subsection{Molecular crystals}
To further test the extrapolating ability of the MBD-ML model, a random set of 200 molecular crystal structures from the recently published OMC25 data set was chosen and recomputed at the PBE0+MBD-NL level of theory. To maximize the diversity of the subset, it was ensured that all 200 crystal structures featured unique molecules.

Despite being trained exclusively on small molecules, the MBD-ML predictions faithfully reproduce the PBE0+MBD-NL reference. The ratios are predicted with a RMSE of $0.02-0.03$, and the resulting MBD energies and forces are accurate to within 1 meV/atom and 1 meV/\AA/atom, respectively, as illustrated in Fig.~\ref{fig:c6-a0-e-f-s-performance-omc25}. While the magnitude of MBD contributions to the stress tensor $\sigma_{\text{MBD}}$ range only between $0$ and $15\,\text{meV/\r{A}}^3$, the MBD-ML model predicts both diagonal and off-diagonal stress components with a very high accuracy, yielding a RMSE of $0.143\,\text{meV/\r{A}}^3$ (see Fig.~\ref{fig:c6-a0-e-f-s-performance-omc25}).

It is important to note that molecular-crystal unit cells often contain more than 100 atoms; consequently, even a modest per-atom error of $0.8$ meV can translate into a total-energy deviation exceeding $80$ meV. However, this consideration only applies to total energies, as such errors cancel out almost exactly for energy differences, resulting in correct polymorph rankings, as shown in the relevant section below. To elucidate the origin of this deviation, we performed a sensitivity analysis of the MBD energy with respect to the $C_6^{\rm r}$ and $\alpha_0^{\rm r}$ ratios. We found that both systematic bias and random noise in these ratios influence the resulting MBD energy, with the effect of bias being approximately twice as large as that of noise. Taken together, these contributions account for the overall deviation of the MBD-ML energy. Details of this analysis are provided in Section~S6 of the Supplementary Information.

Overall, the MBD-ML model demonstrates robust accuracy and transferability across chemical environments. Although trained exclusively on small single-molecule structures from the QCML data set, the predictive accuracy of the model extends to periodic molecular crystals containing multiple molecules of varying complexity.

\begin{figure*}[ht]
    \centering
     \begin{subfigure}[t]{0.3\textwidth}
     \centering
     \caption{}
     \includegraphics[width=\textwidth]{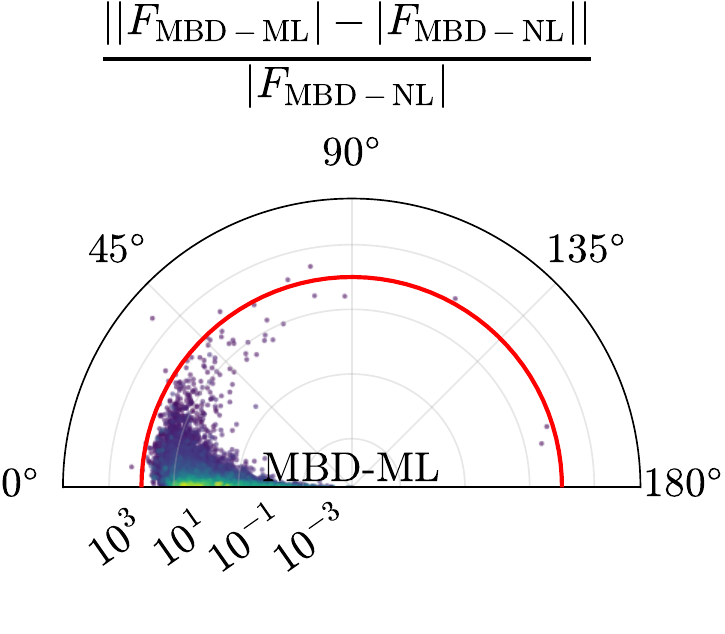}
     \label{fig:mbdml-polar}
     \end{subfigure}
     \begin{subfigure}[t]{0.3\textwidth}
     \centering
     \caption{}
     \includegraphics[width=\textwidth, ,angle=0]{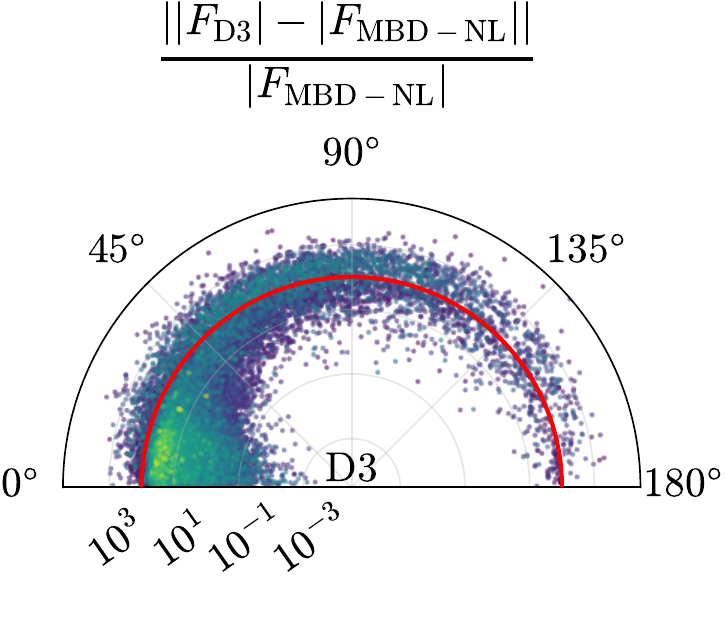}
     \label{fig:d3-polar}
     \end{subfigure}
     \begin{subfigure}[t]{0.365\textwidth}
     \centering
     \caption{}
     \includegraphics[width=\textwidth, ,angle=0]{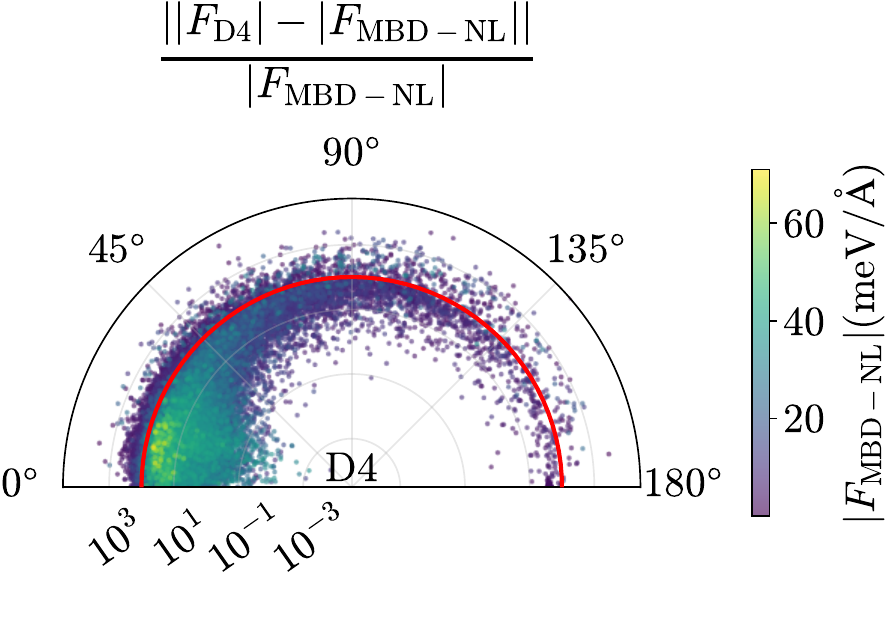}
     \label{fig:d4-polar}
     \end{subfigure}
    \caption{Comparison of MBD-ML and DFT-D performance on predicting
    vdW force contributions compared to MBD-NL.
    Polar representation of atomic force deviations
    computed from the MBD-ML (\textbf{a}), D3 (\textbf{b}) and D4 method
    (\textbf{c}). The radial coordinate specifies the relative magnitude deviation of the atomic force vector to the MBD-NL reference in percent. The angular coordinate corresponds to the angular deviation from the reference in degrees. The point colour encodes the absolute magnitude of the MBD-NL atomic force
    As a guide to the eye the radial $100\%$ mark is highlighted in red}
    \label{fig:overall-mbd-d3-d4-comparison}
\end{figure*}

\section{MBD-ML forces for varying molecule sizes}

Having validated MBD-ML on fixed chemical environments, we now systematically examine whether its force accuracy degrades with increasing molecular size, and benchmark it against the widely used pairwise D3 and D4 corrections.
We selected 825 molecules from the OMol25 dataset~\cite{levine2025open} with sizes uniformly distributed from 3 to 350 atoms. To avoid artifacts, only uncharged, spin-non-polarized molecules were included, excluding systems containing alkali or alkaline earth elements (performance including these rare-in-data elements is reported in Section~S5). Atomic forces were computed at the PBE0+MBD-NL level and compared to predictions from MBD-ML, D3~\cite{grimme2010consistent, grimme2011effect}, and D4~\cite{caldeweyher2017extension}. Fig.~\ref{fig:overall-mbd-d3-d4-comparison} shows magnitude and angular deviations relative to MBD-NL forces, color-coded by the magnitude of MBD-NL force vectors. An in-depth statistical analysis appears in Fig.~S12 and Table~S2.

MBD-ML forces show excellent agreement with MBD-NL references, consistent with QCML and DES370k results. The median and mean magnitude deviations are $2–3\%$ with a $95^{th}$ percentile of $7.7\%$. Angular deviations average $<2^{\circ}$ with a $95^{th}$ percentile of $5.1^{\circ}$. While Fig.~\ref{fig:overall-mbd-d3-d4-comparison} shows a few MBD-ML outliers with angular deviations exceeding $150^{\circ}$, these occur exclusively when reference forces are nearly zero (deep purple), making such deviations practically irrelevant.

In contrast, D3 and D4 are strikingly different, with deviations in magnitude and angle being approximately ten times larger than for MBD-ML -- $28.2\%$ and $17.4^{\circ}$ for D3 and $39.5\%$ and $19.8^{\circ}$ for D4, respectively. This is consistent with previously reported disagreements between MBD and pairwise methods for atomic chains~\cite{hauseux2022colossal}, non-covalent dimers~\cite{puleva2025extending}, and polymers~\cite{sosa2025power}. This qualitative difference also manifests itself in extreme outliers (force magnitude deviations exceeding $100\%$). D3 produces 3066 such outliers ($2.4\%$ of all atoms) and D4 produces 4025 ($3.2\%$), compared to only 8 atoms ($0.006\%$) for MBD-ML.

Interestingly, D4 shows larger deviations from MBD-NL for both force magnitudes and angles than D3. The polar representation in Fig.~\ref{fig:overall-mbd-d3-d4-comparison} reveals an important qualitative difference: while both methods show substantially broader deviation distributions than MBD-ML, in the case of D4 large angular deviations exceeding $65^{\circ}$ predominantly occur for very small force magnitudes (deep purple markers). In contrast, D3 forces exhibit sizeable angular deviations for a higher fraction of large vdW forces (green markers) as well. This distinction is critical, as deviations in larger forces are more consequential for practical applications. 

These results demonstrate that MBD-ML closely reproduces the MBD-NL reference across diverse molecular sizes, while emphasizing the importance of many-body effects beyond pairwise approximations. The substantial deviations of DFT-D methods from MBD-NL, particularly for larger forces where accuracy is most critical, underscore the importance of many-body dispersion effects that pairwise approaches fundamentally cannot capture.

\section{Molecular crystal structure and stability}
Predicting the relative stability of molecular crystal polymorphs is among the most demanding practical tests for a dispersion method, as energy differences between forms often fall below 1 kJ/mol per molecule.
To assess MBD-ML in this context, we optimized geometries and computed energy rankings at the DFT+MBD-ML level of theory for a set of crystals spanning hydrogen-bonded, van der Waals, and mixed interaction types (see Table~S3 and Section~S8).

\begin{figure}
\includegraphics[width=0.99\linewidth]{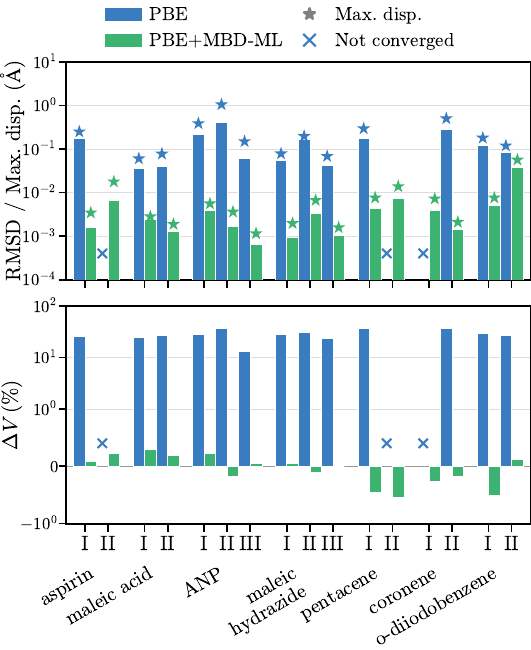}
\caption{Structural comparison of polymorphs relaxed with PBE and PBE+MBD-ML compared to the PBE+MBD-NL relaxed structures: (top) RMSD and maximum atomic displacement and (bottom) ratio of the unit cell volume of the relaxed structures with respect to the PBE+MBD-NL-relaxed structure. Blue crosses represent cases where the PBE geometry optimization did not converge. The RMSD and volume ratio is shown for every polymorph of each molecular crystal. ANP refers to 2-amino-5-nitropyrimidine. Numerical values are tabulated in Table~S4}
\label{fig:rmsd-and-vol-ratio-pbe-vs-pbembdml}
\end{figure}

\subsection{Geometry optimization}
We first assess the quality of the optimized structures before evaluating their energy rankings.
PBE+MBD-ML closely reproduces PBE+MBD-NL~\cite{perdew1996generalized} crystal structures, with RMSDs of atomic positions ranging from $0.001\,\mathrm{\r{A}}$ to $0.05\,\mathrm{\r{A}}$ for most structures (Fig.~\ref{fig:rmsd-and-vol-ratio-pbe-vs-pbembdml}). Unit cell volumes differ by less than $1\%$ from the \emph{ab initio} reference, demonstrating that MBD-ML forces and stresses are sufficiently accurate to yield virtually identical equilibrium geometries. These small deviations are within typical uncertainties of experimental crystal structures and DFT geometry optimizations, confirming MBD-ML's reliability for structural predictions. In contrast, uncorrected PBE substantially overestimates unit cell volumes by over $20\%$ due to missing attractive dispersion interactions, with RMSDs one to two orders of magnitude higher than MBD-ML (Fig.~\ref{fig:rmsd-and-vol-ratio-pbe-vs-pbembdml}). This comparison underscores the necessity of correctly accounting for dispersion interactions in molecular crystals.

\begin{table}
\caption{Polymorph energy differences (kJ/mol per molecule) from PBE0+MBD-NL ($\Delta E_{\mathrm{MBD-NL}}$) and PBE0+MBD-ML ($\Delta E_{\mathrm{MBD-ML}}$). Roman numerals indicate the polymorphic forms between which the energy difference is computed (e.g., I $\rightarrow$ II denotes $E_{\mathrm{II}} - E_{\mathrm{I}}$). Systems are grouped by dominant intermolecular interaction. ANP refers to 2-amino-5-nitropyrimidine. 
Incorrect rankings likely due to incomplete cancellation of total energy errors are in bold.
\textsuperscript{a}Structures were taken from Ref.~\citenum{potticary2016unforeseen}}
\centering
\setlength{\tabcolsep}{4pt}
\begin{tabular}{l l cc}
\toprule
Molecule & & \makecell[c]{$\Delta E_{\mathrm{MBD-NL}}$\\$(\mathrm{kJ/mol})$} & \makecell[c]{$\Delta E_{\mathrm{MBD-ML}}$\\$(\mathrm{kJ/mol})$} \\
\midrule
\multicolumn{4}{l}{\textit{Mixed Interactions}}\\[2pt]
\quad Aspirin & I $\rightarrow$ II & -0.262 & -0.291\\
\midrule
\multicolumn{4}{l}{\textit{Hydrogen Bonded}}\\[2pt]
\quad Maleic acid & I $\rightarrow$ II & 0.670 & 0.467\\
\quad ANP & I $\rightarrow$ II & 0.183 & \textbf{-0.292}\\
\quad & I $\rightarrow$ III & 0.237 & 0.261\\
\quad Maleic hydrazide & I $\rightarrow$ II & 0.974 & 0.870\\
\quad & I $\rightarrow$ III & -0.384 & \textbf{0.220}\\
\midrule
\multicolumn{4}{l}{\textit{van der Waals}}\\[2pt]
\quad Pentacene & I $\rightarrow$ II & -0.476 & -0.622\\
\quad Coronene\textsuperscript{a} & I $\rightarrow$ II & 0.272 & 0.057\\
\quad \textit{o}-Diiodobenzene & I $\rightarrow$ II & 0.029 & 0.015\\
\bottomrule
\end{tabular}
\label{tab:e-ranking-small-novv}
\end{table}

\subsection{Energy ranking}
We evaluate MBD-ML's reliability in predicting the energetic ordering of polymorphs following the established protocol by Hoja \emph{et al.}~\cite{hoja2019reliable}. Energies of structures optimized at the PBE+MBD-NL and PBE+MBD-ML levels were recomputed using PBE0+MBD-NL and PBE0+MBD-ML, respectively, to determine lattice energy differences. Since MBD-ML is designed as a cost-effective replacement for \emph{ab initio} MBD-NL, we use PBE0+MBD-NL as our reference rather than experimental data, which would require free energy calculations with vibrational entropic contributions beyond the scope of this work.

For seven of nine polymorph transitions studied, MBD-ML correctly predicts the energy ranking with errors below  $0.3\,\text{kJ/mol}$  per molecule (Table~\ref{tab:e-ranking-small-novv}). These systems span diverse interaction types -- van der Waals, hydrogen bonding and mixed, demonstrating MBD-ML's robustness across chemical environments. Given that typical polymorph energy differences are on the order of $0-7\,\mathrm{kJ/mol}$~\cite{nyman2015static}, the MBD-ML errors are well within the range required for reliable polymorph screening.

Two cases exhibit larger errors resulting in incorrect rankings: the I$\rightarrow$II transition of 2-amino-5-nitropyrimidine ($0.5\,\text{kJ/mol}$ per molecule) and the I$\rightarrow $III transition of maleic hydrazide ($0.6\,\text{kJ/mol}$ per molecule). These errors are, however, still very subtle and likely a consequence of the cancellation of total energy errors as previously discussed for molecular crystals of the OMC25 data set.

\section{Limitations of the model}
So far, the MBD-ML model has been shown to be an accurate and reliable dispersion method for diverse systems, including organic and inorganic molecules of widely varying sizes, molecular dimers, and organic crystals. 
Two limitations warrant attention, however. Molecular anions with unbound electrons can yield highly unphysical predicted $a_0^{\mathrm{r}}$ and $C_6^{\mathrm{r}}$ ratios, though this reflects the pathological electronic structure of such systems rather than a deficiency of the model itself. More substantively, the model requires further development for systems containing alkali and alkaline earth elements and for inorganic materials. While the former was discussed in the validation results above, we examine the performance of MBD-ML for inorganic materials in more detail here.

In particular, significantly larger $a_0^{\mathrm{r}}$ and $C_6^{\mathrm{r}}$ errors are observed on a random subset of 203 materials from the OMat24 dataset~\cite{barroso2024open}, recomputed at the HSE06+MBD-NL level of theory. As shown in Table~\ref{tab:performance-cs-models}, the mean absolute errors of 0.247 and 0.269 for these ratios are 10–20 times higher than for all other chemical classes examined in this study. These errors are sufficiently large to introduce false negative MBD eigenvalues, preventing calculation of MBD energies and forces for most materials in the OMat24 subset.

This inadequate performance is likely due to the mismatch between the complex and diverse atomic environments in inorganic materials and those represented in the QCML training dataset. By construction, QCML molecules, derived exclusively from organic molecule databases, contain at most 8 non-hydrogen atoms ~\cite{ganscha2025qcml}, whereas inorganic solids consist almost exclusively of non-hydrogen atoms in extended periodic environments, resulting in minimal overlap between the atomic environments present in the two datasets. 
Crystalline silicon offers a simple illustration. Each silicon atom has tetrahedral coordination with four neighbours, yet any QCML molecule is highly unlikely to contain five silicon atoms in a comparable coordination environment.

Improving MBD-ML performance in inorganic solids and systems containing alkali and alkaline earth metals requires including representative subsets of such systems in the training data, which will be pursued in future work.

\section{Computational scaling}
\begin{figure}
    \centering
    \includegraphics[width=0.9\linewidth]{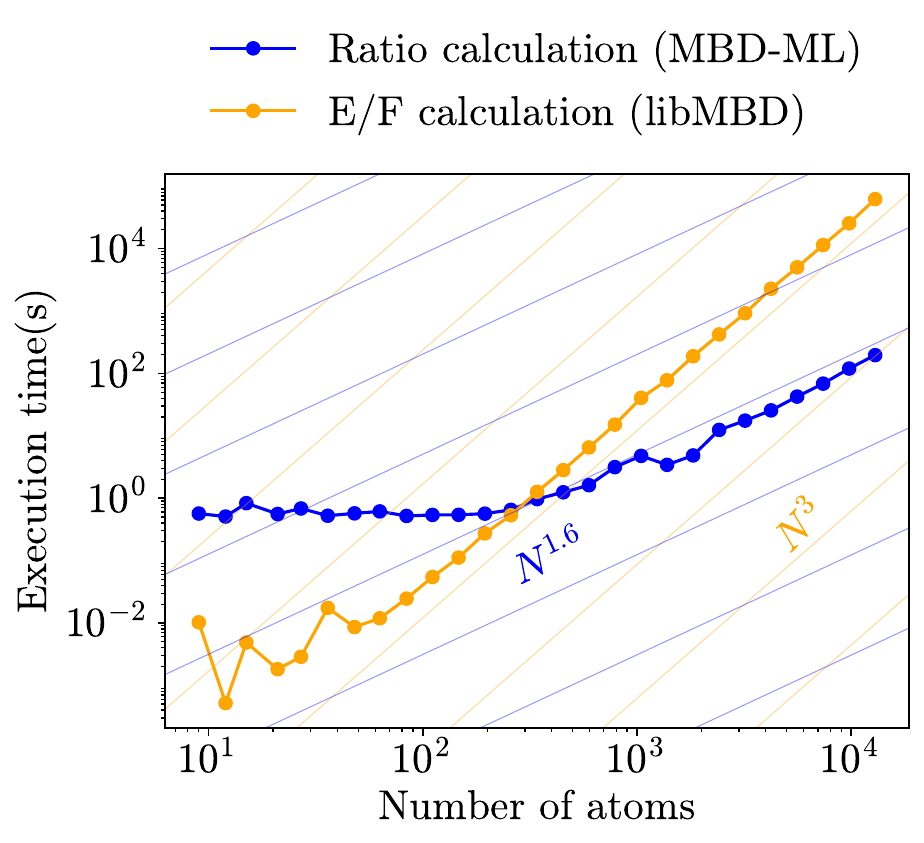}
    \caption{Computational scaling of the calculation of the 
    $a_0^{\mathrm{r}}$ and $C_6^{\mathrm{r}}$ ratio calculation by the MBD-ML model and of the MBD property calculation
    via libMBD}
    \label{fig:comp-scaling}
\end{figure}

Alongside accuracy, computational efficiency is a key requirement for any method intended for large-scale applications. Although the formal scaling of MBD with system size is cubic, its cost in practice remains a small fraction of a single DFT self-consistency cycle. Until now, however, obtaining the $a_0^{\mathrm{r}}$ and $C_6^{\mathrm{r}}$ ratios required a full DFT calculation, making this the true computational bottleneck. MBD-ML removes this bottleneck, reducing the cost of a complete MBD calculation to that of a neural network inference followed by a matrix diagonalization. To quantify the resulting performance, we characterize the scaling of MBD-ML and libMBD as a function of system size.

We computed MBD corrections to total energies and forces for water clusters (H$_2$O)$_{n}$ with $n=3-4321$ (up to nearly 13,000 atoms) serially on a single 128-core node (AMD EPYC Rome); the results are summarized in Fig.~\ref{fig:comp-scaling}. For systems up to 1000 atoms, the MBD-ML calculation time for $a_0^{\mathrm{r}}$ and $C_6^{\mathrm{r}}$ remains approximately constant at only a few seconds. The subsequent MBD energy and force calculations exhibit the theoretically predicted cubic scaling beyond 60 atoms, and are faster than the ratio calculations up to a crossover point at 250 atoms. For clusters exceeding 1000 atoms, the MBD-ML execution time grows with an approximate power of 1.6, reaching a maximum of 196 seconds for the (H$_2$O)$_{4321}$ cluster.

This analysis demonstrates that integrating MBD-ML into the libMBD infrastructure enables MBD calculations of unprecedented scale that would otherwise be impractical with state-of-the-art DFAs.

\section{Summary and Conclusions}
We have developed MBD-ML, a pretrained machine learning model that serves as an accurate and efficient replacement for \textit{ab initio} MBD-NL calculations. By directly predicting polarizability $\alpha_0^{r}$ and $C_6^{r}$ ratios from atomic coordinates, MBD-ML enables computation of MBD energies, forces, and stresses without resorting to any electronic structure calculations. Integration with \texttt{libMBD} reduces the entire workflow to a single function call in the \texttt{pymbd} interface, facilitating incorporation of MBD-ML into a wide range of atomistic simulation workflows, including those built on the Atomic Simulation Environment (\texttt{ASE})~\cite{larsen2017atomic}.

We have demonstrated the reliability, accuracy, and transferability of MBD-ML across diverse systems: single molecules of widely varying sizes and chemical compositions, molecular dimers, and organic crystals. Compared to \textit{ab initio} MBD-NL, MBD-ML reproduces energies, forces, and stresses with errors below $1\,\text{meV/atom}$, $1\,\text{meV/\AA}$, and $0.2\,\text{meV/\AA}^3$, respectively.
Molecular crystal geometry optimizations further confirm this performance, with structural RMSDs of $10^{-3} - 10^{-2}\,\mathrm{\AA}$ and polymorph energy ranking errors of at most $0.6\,\mathrm{kJ/mol}$.

This development addresses a practical barrier to the wider adoption of many-body dispersion methods, namely that the accuracy of MBD-NL has until now come at the cost of requiring a full electronic structure calculation. Early implementations in codes such as VASP and Quantum~\textsc{Espresso} suffered from computational inefficiencies~\cite{hermann2023libmbd}; the libMBD library~\cite{hermann2023libmbd} largely resolved these, and MBD-ML now removes the remaining barrier by decoupling many-body dispersion calculations from any underlying electronic structure code. Beyond facilitating MBD-NL adoption in quantum chemistry codes, MBD-ML also enables more accurate dispersion treatments in machine learning force fields~\cite{kabylda2025molecular}.

Several limitations have been identified for future development. 
The model's applicability to inorganic solids and systems containing alkali or alkaline earth metals is currently limited by under-representation of these systems in the training data, and can in principle be improved by dataset expansion and fine-tuning. The treatment of molecular anions poses a more fundamental challenge, as accurately describing the electronic structure of negatively charged molecules is a well-known difficulty that, as demonstrated in this work, is further exacerbated within the MBD formalism. Generating reliable training data for such systems will require additional work to identify suitable basis set settings and density functional approximations.

\vspace{-2mm}
\section*{Code Availability}
The MBD-ML model has been integrated into the \textsc{Python}
interface of the libMBD library, \texttt{pymbd} (\url{https://github.com/em819/libmbd.git}). All \emph{ab initio}
electronic structure calculations have been 
performed with \textsc{FHI-aims}. For the geometry optimizations,
the Atomic Simulation Environment (ASE) interface~\cite{larsen2017atomic} to
\textsc{FHI-aims} was employed. RMSDs were computed using the \texttt{pymatgen}~\cite{ong2013python} package.

\vspace{-2mm}
\section{Acknowledgements}
All calculations were performed on the Luxembourg national supercomputer MeluXina.

\vspace{-2mm}
\section{Funding}
E.M., A.K., A.Kh., and A.T. acknowledge the European Research Council under ERC-AdG grant FITMOL (101054629). A.T. acknowledge the Luxembourg National Research Fund under grant FNR-CORE MBD-in-BMD (18093472). A.K. acknowledges financial support from the Luxembourg National Research Fund (FNR AFR Ph.D. Grant 15720828).

\vspace{-2mm}
\section{Contributions}
A.T. conceived and supervised the project. A.K. trained the ML model. E.M and A.Kh. integrated the MBD-ML model into the libMBD library. E.M. performed all calculations and data analysis of the main results and drafted the manuscript with input from all authors. A.Kh. analysed the challenges involving molecular anions. E.M. and A.K. created the figures. All authors discussed the results and contributed to editing the manuscript.

\twocolumngrid

\bibliography{references}

\clearpage
\renewcommand{\thesection}{S\arabic{section}}  
\renewcommand{\thetable}{S\arabic{table}}  
\renewcommand{\thefigure}{S\arabic{figure}}
\setcounter{figure}{0}
\setcounter{table}{0}
\onecolumngrid
\end{document}


\title{Supplementary Information for: \\ MBD-ML: Many-body dispersion from machine learning for \\ molecules and molecular crystals}

\maketitle

\section{MBD-ML training and parameters}
The MBD-ML model is based on the SO3LR~\cite{kabylda2025molecular} architecture,
with its long-range contributions disabled, so that
the remaining model is equivalent to SO3krates.
For the purpose of the MBD-ML model, the SO3LR architecture was
extended to also predict $C_6^{r}$ and $\alpha_0^{r}$ ratios.
The training was conducted on the QCML data set for which a combined loss function including these ratios with equal weighting was employed, using a robust loss function with $\alpha = 1$. The learning rate was adapted exponentially using the AMSGrad optimizer with a learning rate of $0.001$ and a decay rate of $0.9$ applied every 1M steps. 
Compared to the original SO3LR model, MBD-ML was kept relatively light weight, with a feature dimension of $64$, 4 heads, a maximal degree of $4$ and $32$ radial basis functions.
Thus, with long-range contributions disabled and a short-range cutoff radius of 
$4.0\,\text{\AA}$, MBD-ML exhibits an effective receptive field of $8.0\,
\text{\AA}$.

\section{Selection of data subsets}
\textbf{DES370k}
We used the dimer structures of the DES370k set, which were recomputed on the
PBE0+MBD-NL level of theory as part of the recently published QCell biomolecular data set~\cite{kabylda2025qcell}.

\textbf{OMol25}
For the construction of the OMol25 subset used in this study, the structures of the SPICE and the PDB protein pocket fragments  subset as the union of those two
contains molecules ranging from 3 to 350 atoms. To achieve a uniform selection
of system sizes, the molecules of these two sets were binned by the number
of atoms, using bins of size 10 ($1-10$ atoms, $11-20$ etc.) and a total of 
883 molecules was randomly picked in a uniform fashion.

\textbf{OMC25}
To ensure a maximally diverse subset for the present work, each of the 
200 molecular crystals that were picked feature a unique molecule, i.e
there are no two polymorphs of the same molecular crystal. Other than that
the structures were chosen randomly without additional constraints.

\textbf{OMat24}
Starting from the OMat24 validation set, 6 materials categories were defined
based on the band gap and the number of atoms in the unit cell:
\begin{enumerate}
    \item \texttt{small-metal}
    \item \texttt{small-insulator}
    \item \texttt{intermediate-metal}
    \item \texttt{intermediate-insulator}
    \item \texttt{large-metal}
    \item \texttt{large-insulator},
\end{enumerate}

where \texttt{metal} and \texttt{insulator} structures were defined
by the PBE band gap of the original OMat24 entry to be less than $0.5\,\text{eV}$ and $\geq 0.5\,\text{eV}$, respectively, while \texttt{small}, 
\texttt{intermediate} and \texttt{large} structures were made up of at most
50 atoms, at most 100 atoms and over 100 atoms, respectively.
For each of these 6 categories, up to 50 materials were randomly selected and
recomputed on the HSE06+MBD-NL level of theory, of which 203 calculations
were completed succesfully and used to benchmark the MBD-ML model in this work.

\section{\emph{Ab initio} calculations}
All DFT re-calculations of the OMol25, DES370k, OMC25 and OMat24 data sets were performed using the FHI-aims electronic
structure package~\cite{blum2009ab, abbott2025roadmap} (version 231212.1). All molecules and molecular crystal
structures, including the subsets of the OMol25, DES370k and OMC25 data sets
were computed on the PBE0+MBD-NL level of theory, while the inorganic material
structures from the OMat24 data set were computed on the HSE06+MBD-NL level.
For the convergence of the SCF-cycle the default accuracy settings in combination
with default tight basis sets were used. Relativistic effects were included
using the scalar atomic ZORA method. For both hybrid functionals 
a MBD damping parameter $\beta$ of $0.83$ was used.
For all molecular crystals, k-meshes with a spacing of $0.03\,\text{\AA}^{-1}$
were employed after finding that for a random subset of 50 OMC25 molecular
crystal structures this yielded forces converged to below $1\,\text{meV/\AA}$.
For the inorganic materials, the k-mesh was chosen
based on the PBE band gap provided in the original OMat24 data set. For 
materials with a PBE band gap of less than $0.5\,\text{eV}$, a very fine k-mesh
spacing of $0.0125\,\text{\AA}^{-1}$ was employed, while materials with
larger band gaps were recomputed on the HSE06+MBD-NL level of theory 
with a coarser k-spacing of $0.02\,\text{\AA}^{-1}$ was used.

\section{Exclusion of anions from QCML}\label{sec:negative-charges}
\subsection{Basis set convergence for anions}
For the QCML subset of negatively charged molecules, we identified outliers in the ratios $\alpha_0^r$ and $C_6^r$ in the original data. These outliers correspond to atoms with unusually large values, reaching 20–60, whereas both ratios typically lie within the 0–2 range. In most cases, the anomalous values occur for hydrogen atoms, which is particularly unexpected.

A detailed analysis of the underlying \textit{ab initio} calculations revealed that, for anions, $\alpha_0^r$ and $C_6^r$ are highly sensitive to the \texttt{cut\_pot} parameter in FHI-aims. This parameter defines the onset of a steeply increasing confining potential $v_{\rm cut}(r)$ used to localize numerical atom-centered radial functions. For more details on $v_{\rm cut}(r)$ and its role in constructing basis functions, we refer the reader to Section 3 of the original FHI-aims paper~\cite{blum2009ab}. 

In practical terms, \texttt{cut\_pot}, together with the width parameter \texttt{w} (kept fixed here), determines the spatial support of a basis function. The default \texttt{tight} settings for organic elements (\texttt{cut\_pot} = 4 \AA, \texttt{w} = 2 \AA) yield a total radial extent of 6 \AA, which is generally sufficient to obtain meV-accurate total energies. However, the accurate description of diffuse electron density in anions requires explicit convergence tests with respect to \texttt{cut\_pot}.

\begin{figure}[H]
    \centering
     \includegraphics[width=0.99\linewidth]{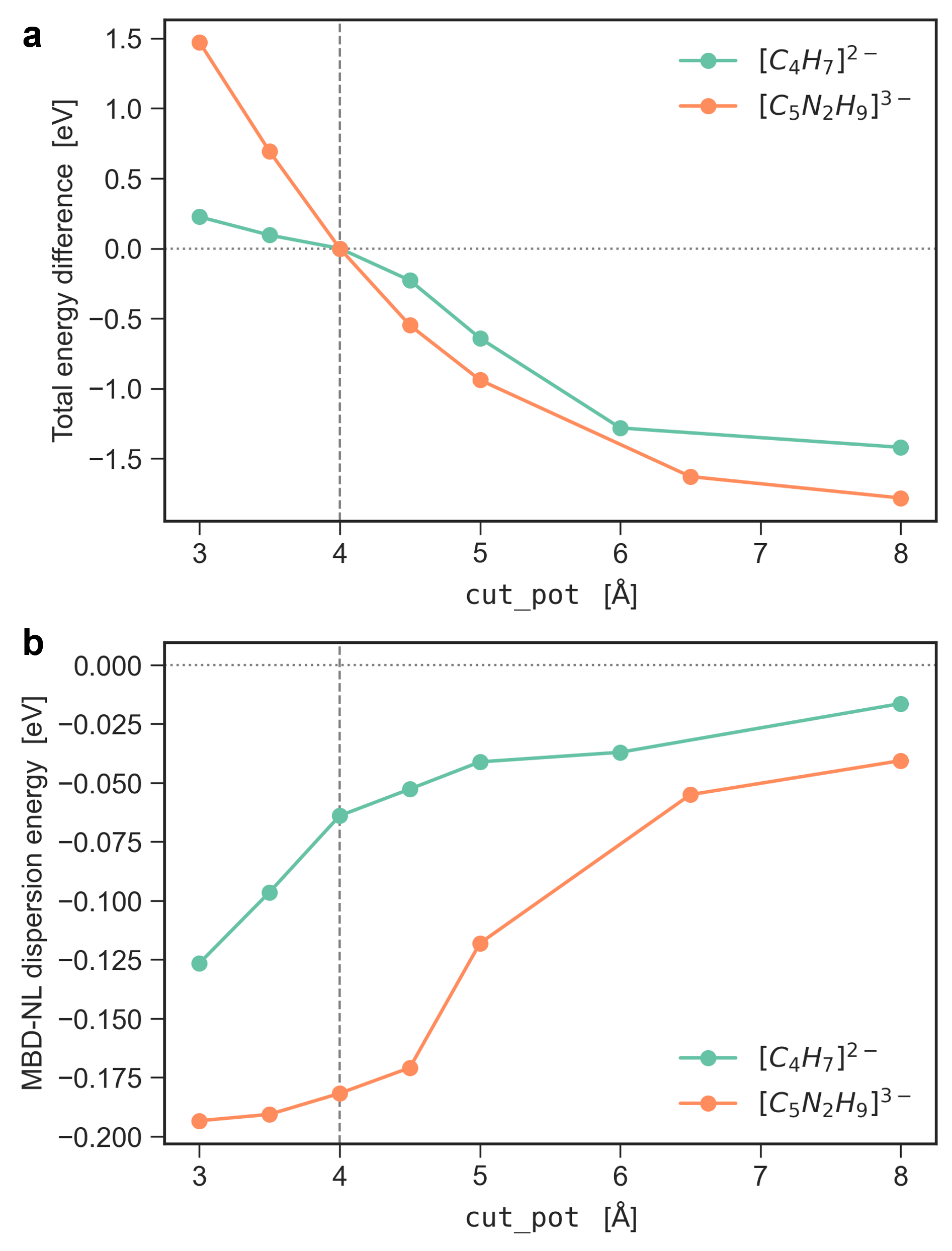}
     \caption{Convergence of ({\textbf a}) total energies and ({\textbf b}) MBD-NL dispersion energies with respect to the \texttt{cut\_pot} parameter for two example anions. Total energies are plotted as the differences relative to the default 4 \AA~setting (marked by a vertical dashed line).}
     \label{fig:anions_en}
\end{figure}

We randomly selected several anions exhibiting anomalous ratios and performed calculations varying \texttt{cut\_pot} between 3 and 8 \AA, while keeping all other computational parameters consistent with QCML settings~\cite{ganscha2025qcml}. The total energies of the anions show strong sensitivity to \texttt{cut\_pot}, varying by up to $\pm 1.5$ eV relative to the default 4 \AA~setting (Figure~\ref{fig:anions_en}a). Convergence is slow, indicating significant contributions from slowly decaying density tails. By contrast, neutral molecules of comparable size (benzene, pentane) exhibit variations of only about 2 meV between \texttt{cut\_pot} = 4 \AA~and 8 \AA.

\begin{figure*}[h!]
    \centering
     \includegraphics[width=0.8\linewidth]{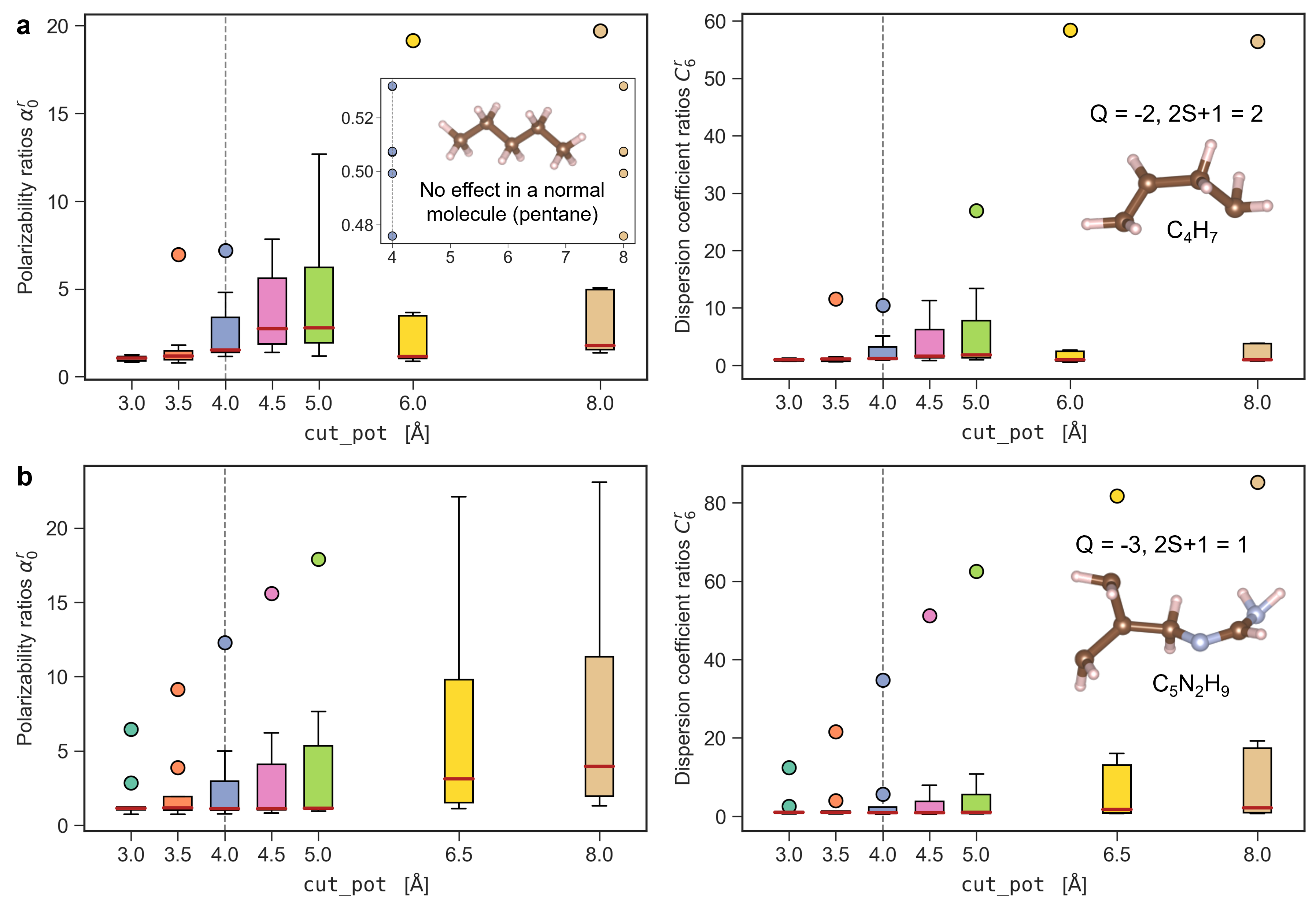}
     \caption{Boxplots of $\alpha_0^r$ (left) and $C_6^r$ (right) with varying  \texttt{cut\_pot} for hydrogen atoms in the two negatively charged molecules: ({\textbf a}) $[\rm C_4H_7]^{2-}$ and ({\textbf b}) $[\rm C_5N_2H_9]^{3-}$ (the structures shown in the insets). The inset in the upper left panel illustrates the $\alpha_0^r$ ratios in pentane molecule for \texttt{cut\_pot} = 4 \AA~and 8~\AA.}
     \label{fig:anions_ratios1}
\end{figure*}

The corresponding MBD-NL dispersion energies are likewise unconverged and increase unphysically as the basis becomes more extended (Figure~\ref{fig:anions_en}b). This behavior is further illustrated by plotting $\alpha_0^{\rm r}$ and $C_6^{\rm r}$ for hydrogen atoms versus \texttt{cut\_pot}, which reveals a dramatic increase in both ratios with increasing basis extent (Figure~\ref{fig:anions_ratios1}). Hydrogen atoms are emphasized because they are typically located at the molecular periphery and thus most sensitive to density tails, although similar trends are observed for some heavier atoms.

Physically, this divergence arises because a more extended basis permits slower decay of the electron density, and the VV polarizability functional is highly sensitive to density gradients in the tail region~\cite{hermann2020density}. The performance of VV and MBD-NL for anions was not assessed in the original work, and this issue has not previously been reported.

Further analysis shows that the problematic species are actually thermodynamically unstable, as indicated by their computed electron affinities (Table~\ref{tab:EA}). For an $N$-electron system, the electron affinity is defined as ${\rm EA}(N) = E(N) - E(N+1)$, with positive values indicating stability of the $(N+1)$-electron state. We evaluated EAs by sequentially removing electrons from the charge states present in QCML. Calculations employed the long-range corrected LC-$\omega$PBEh functional ($\omega = 0.2~{\rm Bohr}^{-1}$)~\cite{rohrdanz2009long} as implemented in \textsc{Q-Chem}~\cite{epifanovsky2021software}, together with aug-cc-pVQZ basis sets to ensure adequate diffuseness~\cite{Kendall1992}.

As summarized in Table~\ref{tab:EA}, only singly charged anions are stable in all four cases; higher charge states are clearly unstable. Consequently, such pathological entries must be recomputed with physically meaningful charge states to form a reliable subset of negatively charged molecules within QCML. To mitigate similar issues in future datasets, additional screening could be applied, for example using HOMO energies as a proxy for stability. As shown in Table~\ref{tab:EA}, HOMO values obtained with a long-range corrected functional correlate well with EAs, consistent with the ionization potential theorem applied to the $(N+1)$-electron system, which yields ${\rm EA}(N) = {\rm IP}(N+1) = -{\rm HOMO}(N+1)$ for the exact functional~\cite{perdew1982}.

A more comprehensive investigation, including systematic benchmarking of basis sets and exchange–correlation functionals, is warranted but lies beyond the scope of the present work. Until such validation and filtering are performed, we do not recommend using QCML data for negatively charged molecules, particularly dispersion energies and derived ratios.

\begin{table}[h!]
\centering
\caption{Total energies, electron affinities and HOMO levels (all in eV) for various charged states of the four considered molecules. The first row in each case corresponds to the charged state included in QCML.} \label{tab:EA}
\begin{tabular}{|c|c|c|c||c|c|c|c|}
\hline
\multicolumn{4}{|c||}{C$_4$H$_7$}                 & \multicolumn{4}{c|}{C$_2$NH$_4$}                 \\
\hline
$Q$ & $E_{\rm tot}$ & EA   & HOMO     & $Q$ & $E_{\rm tot}$ & EA   & HOMO     \\
\hline
-2     & -4252.28   & -2.81  & 2.79  & -2        & -3621.08   & -3.36  & 3.29  \\
\textbf{-1} & \textbf{-4255.09}   & \textbf{0.96}  & \textbf{-1.00} & \textbf{-1}    & \textbf{-3624.44}   & \textbf{1.61}   & \textbf{-1.58} \\
0         & -4254.13   &        & -9.15 &  0        & -3622.84   &        & -9.07 \\
\hline
\hline
\multicolumn{4}{|c||}{C$_5$N$_2$H$_9$}            & \multicolumn{4}{c|}{C$_6$O$_2$H$_9$}             \\
\hline
$Q$ & $E_{\rm tot}$ & EA   & HOMO     & $Q$ & $E_{\rm tot}$ & EA   & HOMO     \\
\hline
-3        & -8290.99   & -4.16 & 4.76  & -3        & -10441.5   & -4.74 & 4.75  \\
-2        & -8295.15   & -2.86 & 2.42  & -2        & -10446.2   & -2.62 & 2.71  \\
\textbf{-1}  & -8298.02   & \textbf{1.67}  & \textbf{-1.72} & \textbf{-1}   & \textbf{-10448.9}   & \textbf{1.30}  & \textbf{-1.82} \\
0         & -8296.35   &       & -7.04 &  0        & -10447.6   &       & -8.77 \\
\hline
\end{tabular}
\end{table}

\begin{figure*}[h!]
    \centering
     \includegraphics[width=0.8\linewidth]{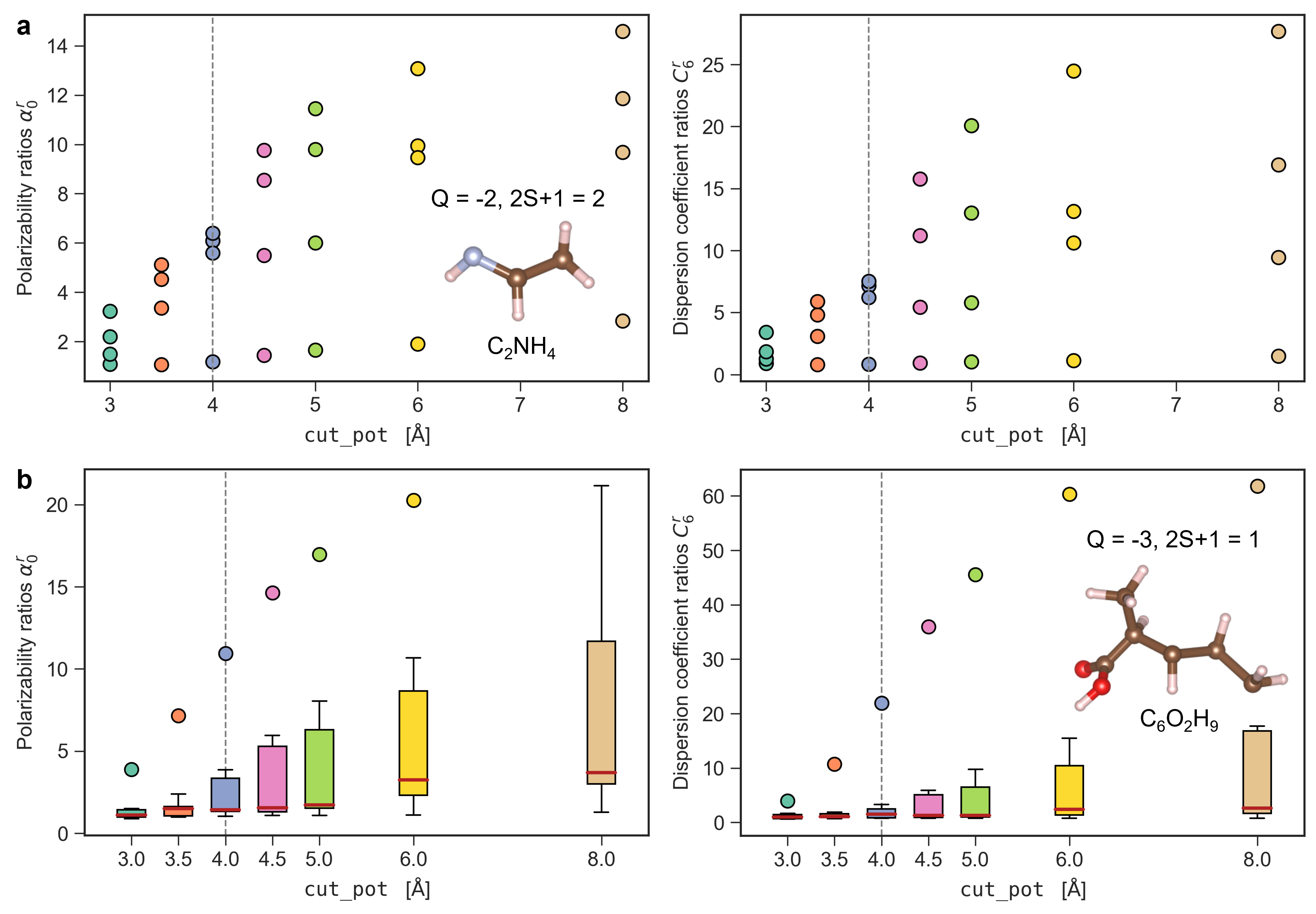}
     \caption{The same as Figure~\ref{fig:anions_ratios1} but for ({\textbf a}) $[\rm C_2NH_4]^{2-}$ and ({\textbf b}) $[\rm C_6O_2H_9]^{3-}$ (the structures shown in the insets). }
     \label{fig:anions_ratios2}
\end{figure*}

\clearpage

\subsection{Performance of MBD-ML with anions included}
\begin{figure}[H]
    \begin{subfigure}[t]{0.49\linewidth}
        \centering
        \caption{}
         \includegraphics[width=\textwidth]{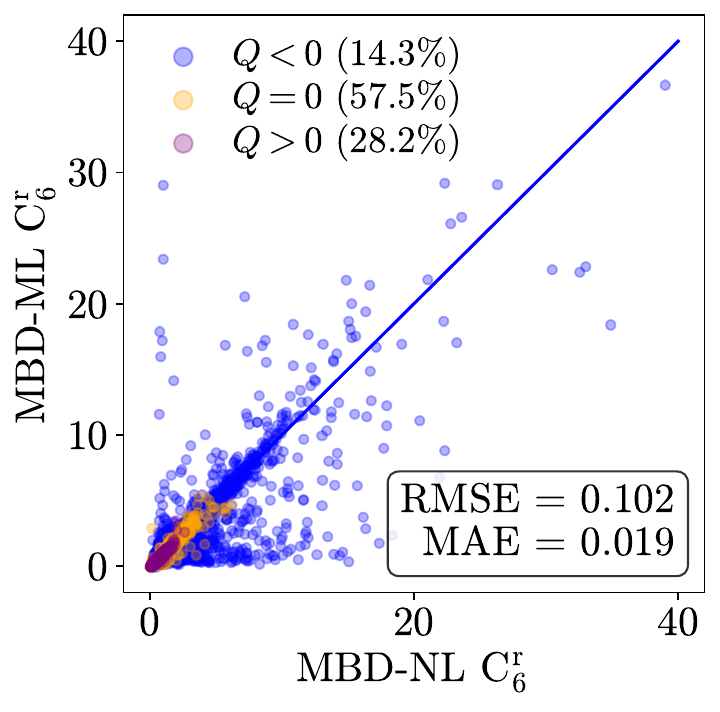}
         \label{fig:mbd-c6-qcml}
    \end{subfigure}
    \begin{subfigure}[t]{0.49\linewidth}
        \centering
        \caption{}
         \includegraphics[width=\textwidth]{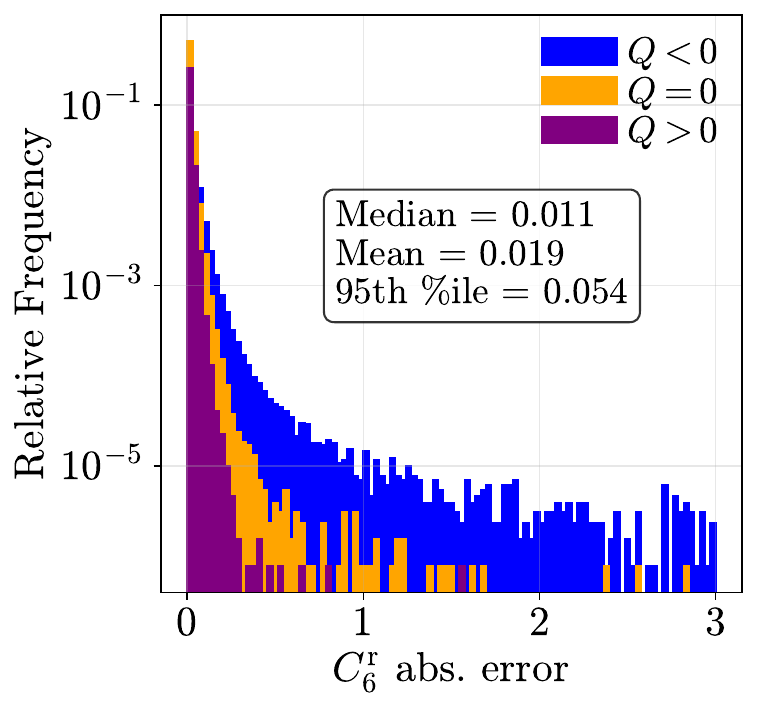}
         \label{fig:mbd-c6-hist-qcml}
    \end{subfigure}
    
    \begin{subfigure}[t]{0.49\linewidth}
        \centering
        \caption{}
        \includegraphics[width=\textwidth]{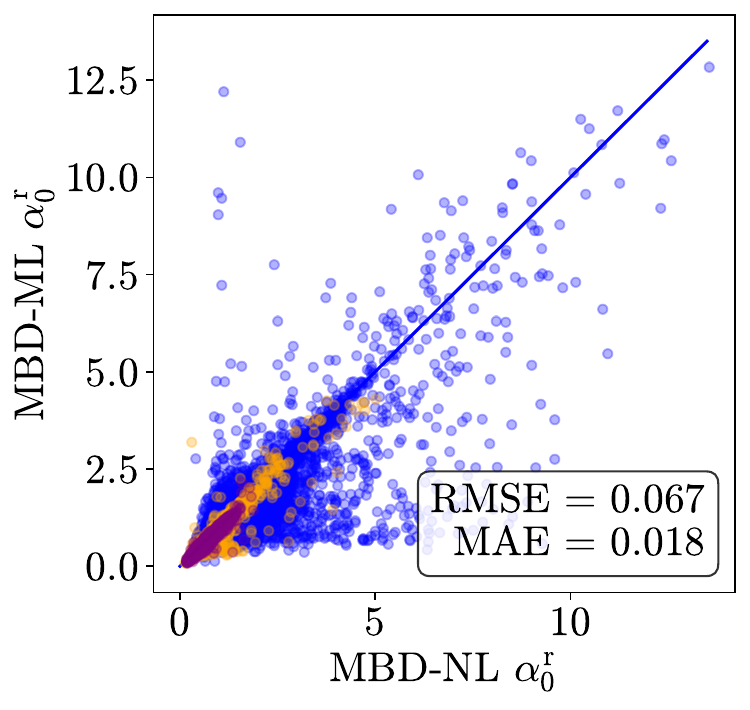}
        \label{fig:mbd-a0-qcml}
    \end{subfigure}
    \begin{subfigure}[t]{0.49\linewidth}
        \centering
        \caption{}
        \includegraphics[width=\textwidth]{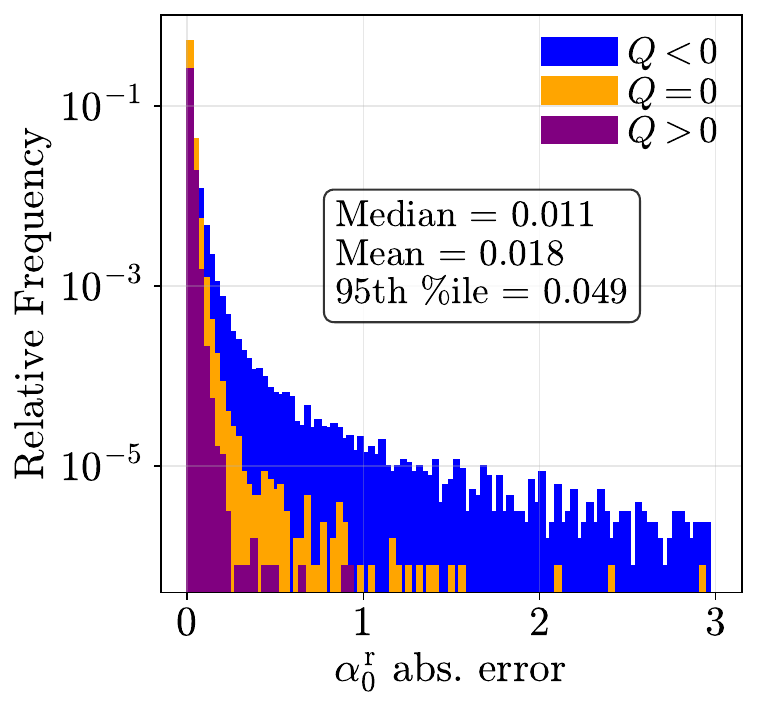}
        \label{fig:mbd-a0-hist-qcml}
    \end{subfigure}
    \caption{Performance of PBE0+MBD-ML in predicting the $C_6$
    and $\alpha_0$ ratios of the QCML test set including all negatively charged molecules}
    \label{fig:c6-a0-performance-qcml-old}
    \end{figure}

If these pathological anions from the QCML data set are included in the accuracy measurement of the MBD-ML model, one finds numerous outliers in the MBD ratio prediction (see Figures~\ref{fig:c6-a0-performance-qcml-old}), which is plausible due to the insufficient computational settings for negatively charged molecules outlined in the previous subsection. 

\begin{figure}[t]
    \begin{subfigure}[]{0.48\linewidth}
        \centering
        \caption{}
        \includegraphics[width=\textwidth]{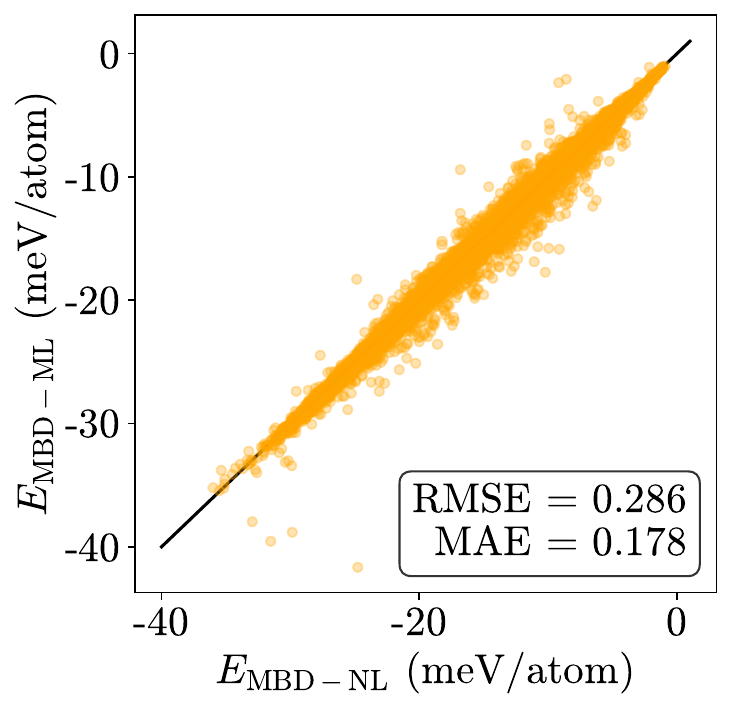}
        \label{fig:mbd-e-qcml}
    \end{subfigure}   
    \begin{subfigure}[]{0.48\linewidth}
        \centering
        \caption{}
        \includegraphics[width=\textwidth]{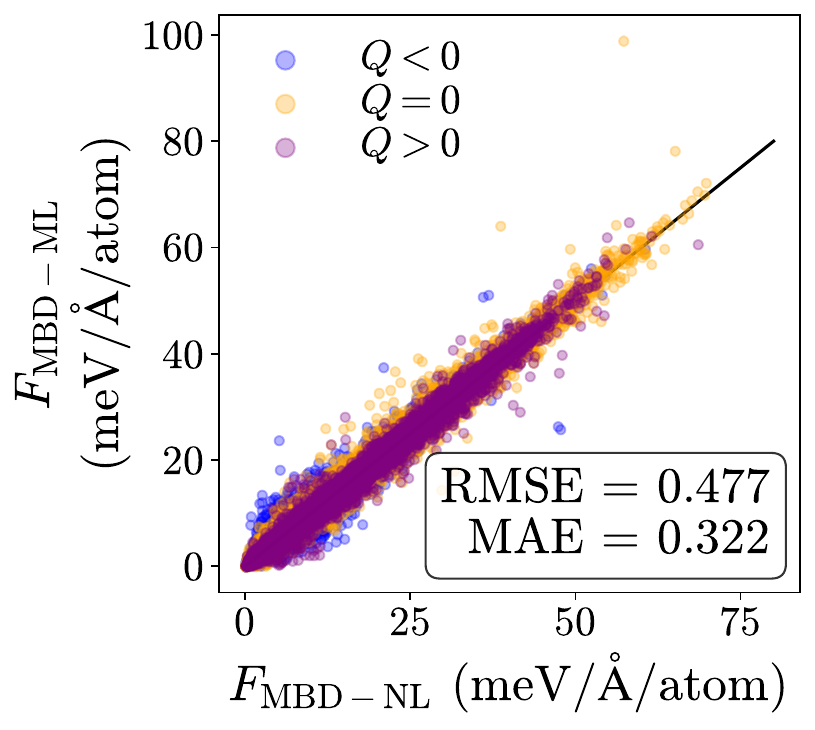}
        \label{fig:mbd-f-qcml}
    \end{subfigure}
    \caption{Performance of PBE0+MBD-ML in predicting the MBD contribution to the total energy and atomic forces of the QCML test set including all negatively charged molecules}
    \label{fig:e-f-performance-qcml-old}
    \end{figure}

We find, however, that the resulting MBD energy and force contributions are basically
unaffected from these outliers (see Figure~\ref{fig:e-f-performance-qcml-old}).

\section{MBD-ML vs MBD-NL performance for OMol25 and DES370k subsets}
\begin{figure}
\centering
\begin{subfigure}{0.48\linewidth}
\caption{}
\includegraphics[width=\linewidth]{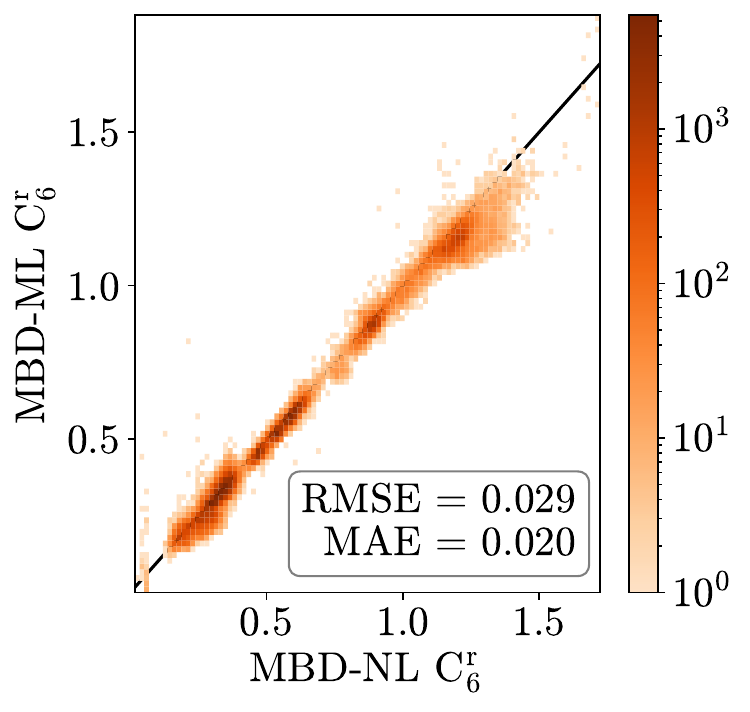}
\end{subfigure}%
\begin{subfigure}{0.48\linewidth}
\caption{}
\includegraphics[width=\linewidth]{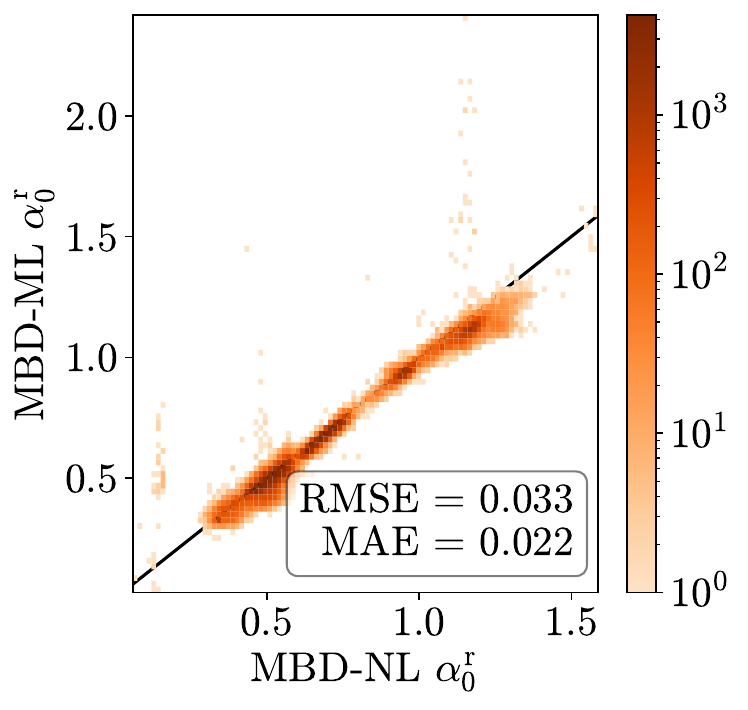}
\end{subfigure}%

\begin{subfigure}{0.48\linewidth}
\caption{}
\includegraphics[width=\linewidth]{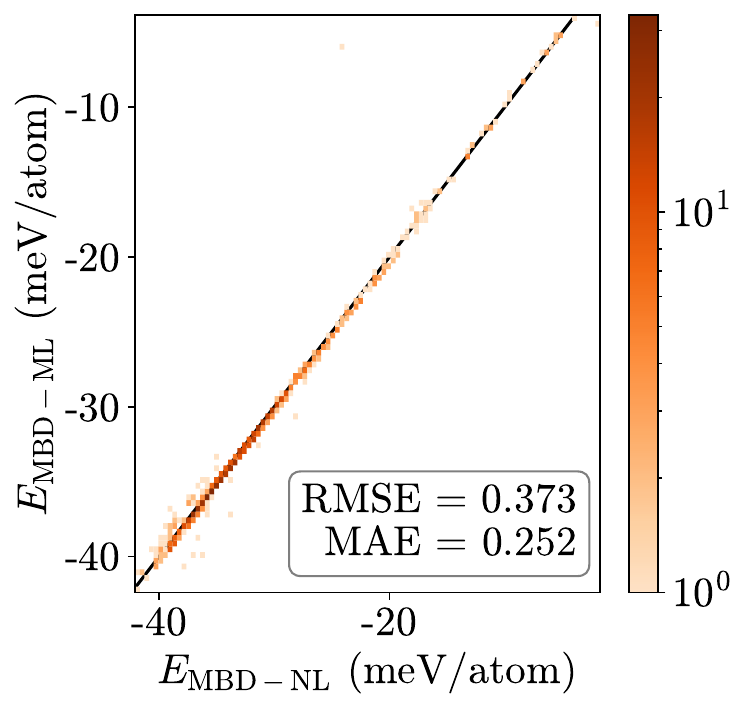}
\end{subfigure}%
\begin{subfigure}{0.48\linewidth}
\caption{}
\includegraphics[width=\linewidth]{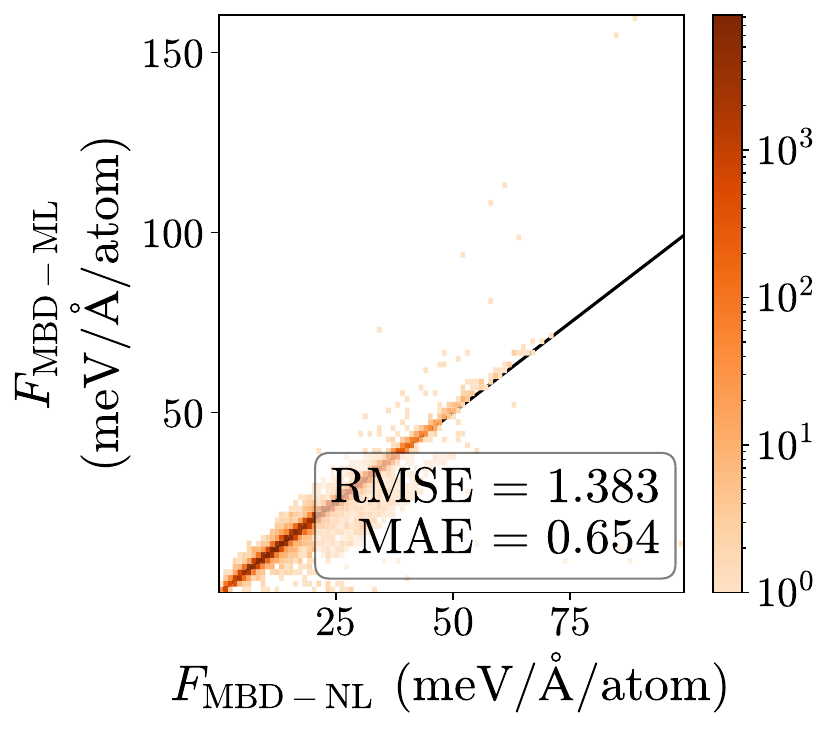}
\end{subfigure}
\caption{Performance of PBE0+MBD-ML in predicting the $C_6$
and $\alpha_0$ ratios and the MBD contribution to the total energy and atomic forces of the OMol25 test set including alkali and alkaline earth elements}
\label{fig:c6-a0-e-f-performance-omol25}
\end{figure}

\begin{figure}
\centering
\begin{subfigure}{0.48\linewidth}
\caption{}
\includegraphics[width=\linewidth]{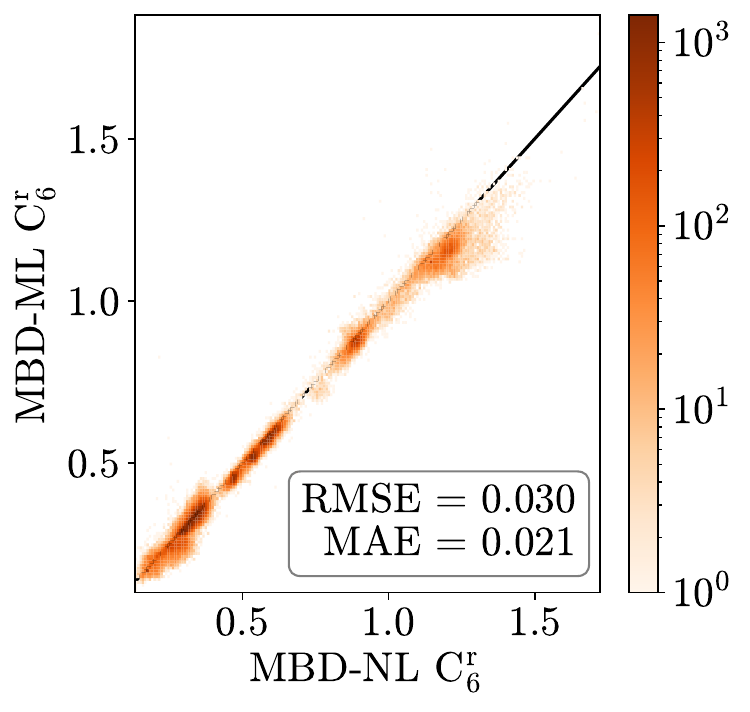}
\end{subfigure}%
\begin{subfigure}{0.48\linewidth}
\caption{}
\includegraphics[width=\linewidth]{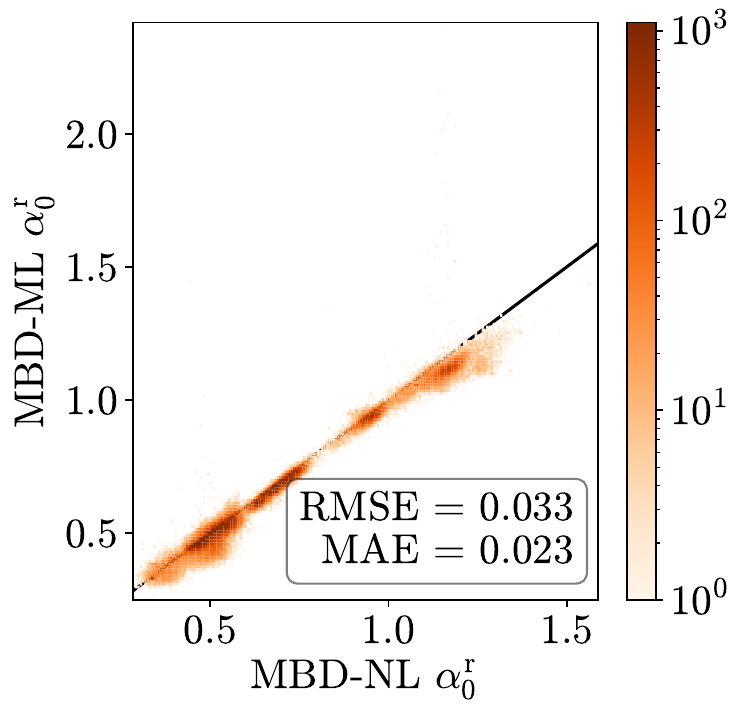}
\end{subfigure}%

\begin{subfigure}{0.48\linewidth}
\caption{}
\includegraphics[width=\linewidth]{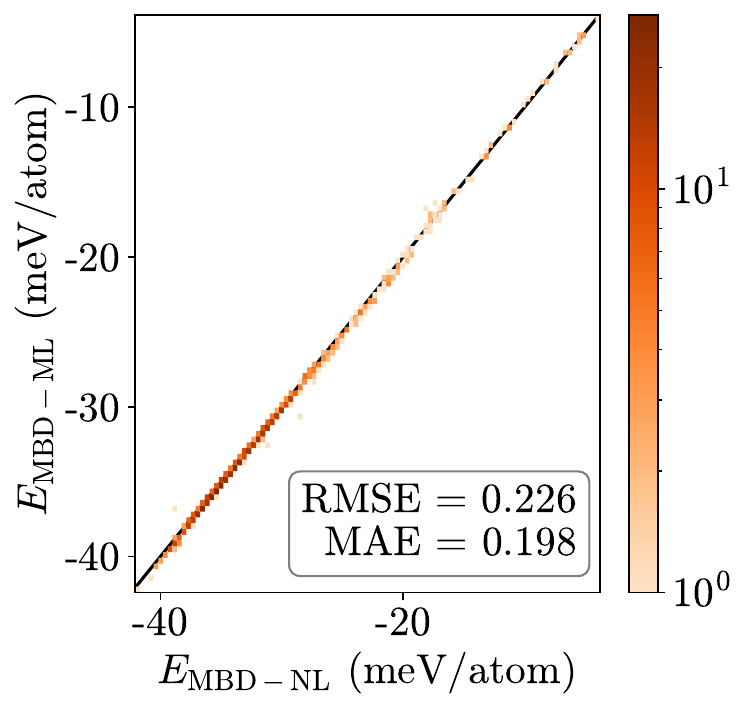}
\end{subfigure}%
\begin{subfigure}{0.48\linewidth}
\caption{}
\includegraphics[width=\linewidth]{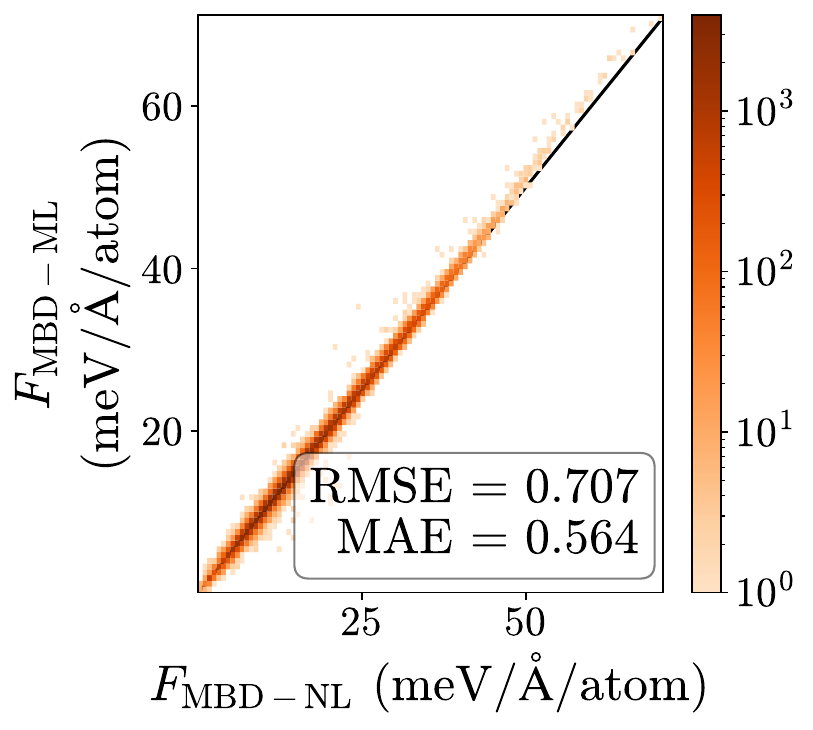}
\end{subfigure}
\caption{Performance of PBE0+MBD-ML in predicting the $C_6$
and $\alpha_0$ ratios and the MBD contribution to the total energy and atomic forces of the OMol25 test set excluding alkali and alkaline earth elements}
\label{fig:c6-a0-e-f-performance-omol25-no-alkali}
\end{figure}

\begin{figure}
\centering
\begin{subfigure}{0.48\linewidth}
\caption{}
\includegraphics[width=\linewidth]{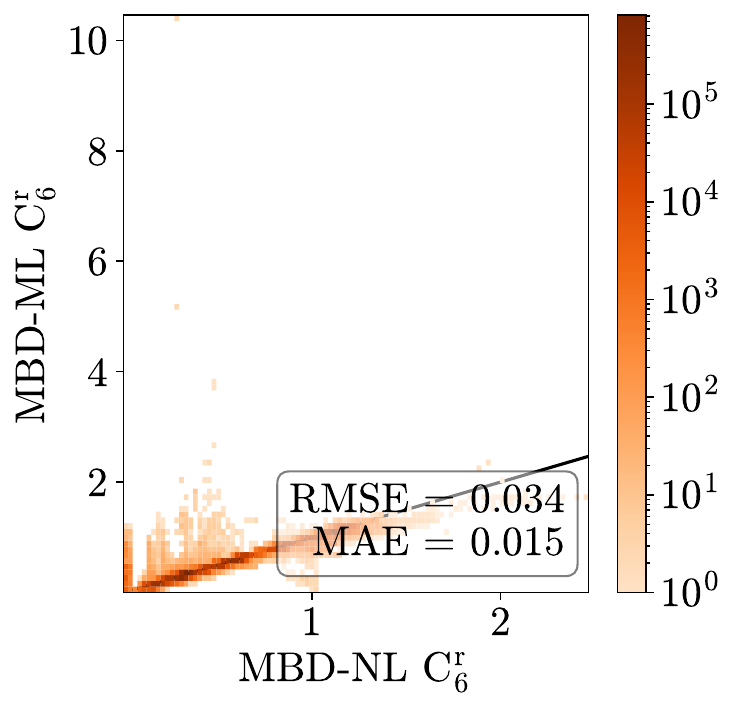}
\end{subfigure}%
\begin{subfigure}{0.48\linewidth}
\caption{}
\includegraphics[width=\linewidth]{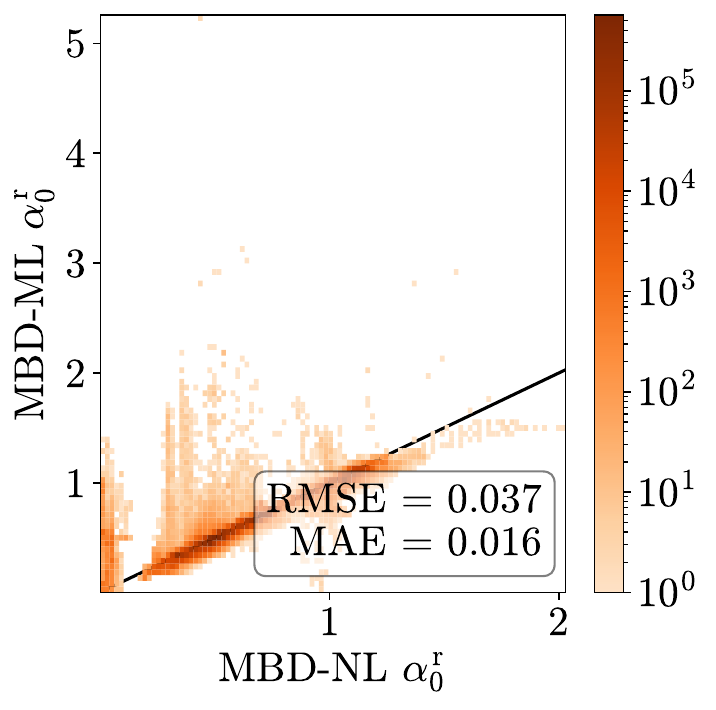}
\end{subfigure}%

\begin{subfigure}{0.48\linewidth}
\caption{}
\includegraphics[width=\linewidth]{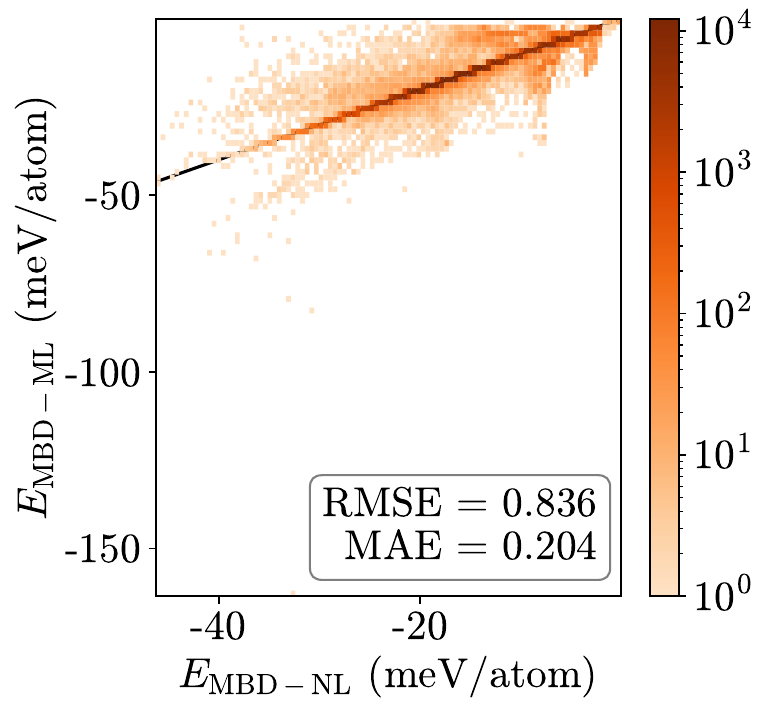}
\end{subfigure}%
\begin{subfigure}{0.48\linewidth}
\caption{}
\includegraphics[width=\linewidth]{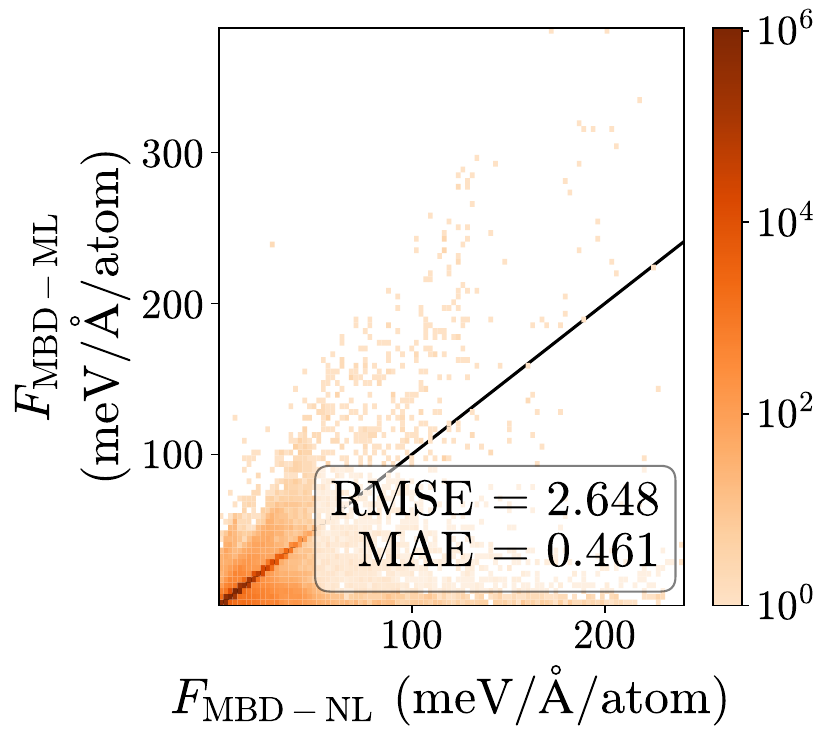}
\end{subfigure}
\caption{Performance of PBE0+MBD-ML in predicting the $C_6$
and $\alpha_0$ ratios and the MBD contribution to the total energy and atomic forces of the DES370k test set including alkali and alkaline earth elements}
\label{fig:c6-a0-e-f-performance-des370k}
\end{figure}

\begin{figure}
\centering
\begin{subfigure}{0.48\linewidth}
\caption{}
\includegraphics[width=\linewidth]{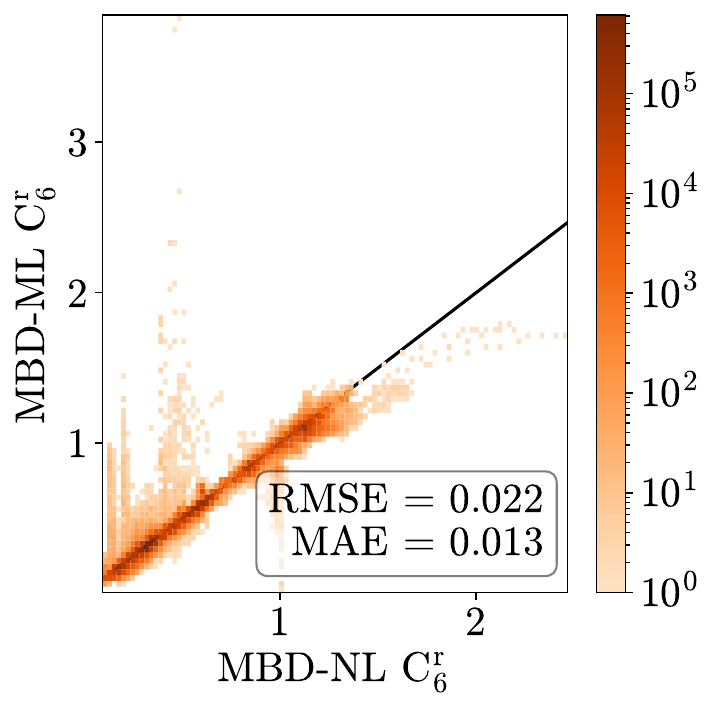}
\end{subfigure}%
\begin{subfigure}{0.48\linewidth}
\caption{}
\includegraphics[width=\linewidth]{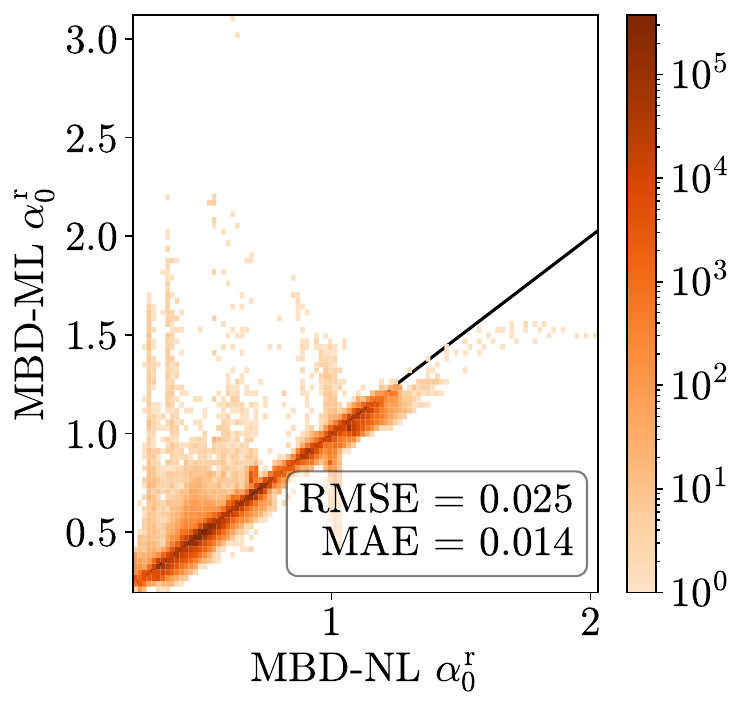}
\end{subfigure}%

\begin{subfigure}{0.48\linewidth}
\caption{}
\includegraphics[width=\linewidth]{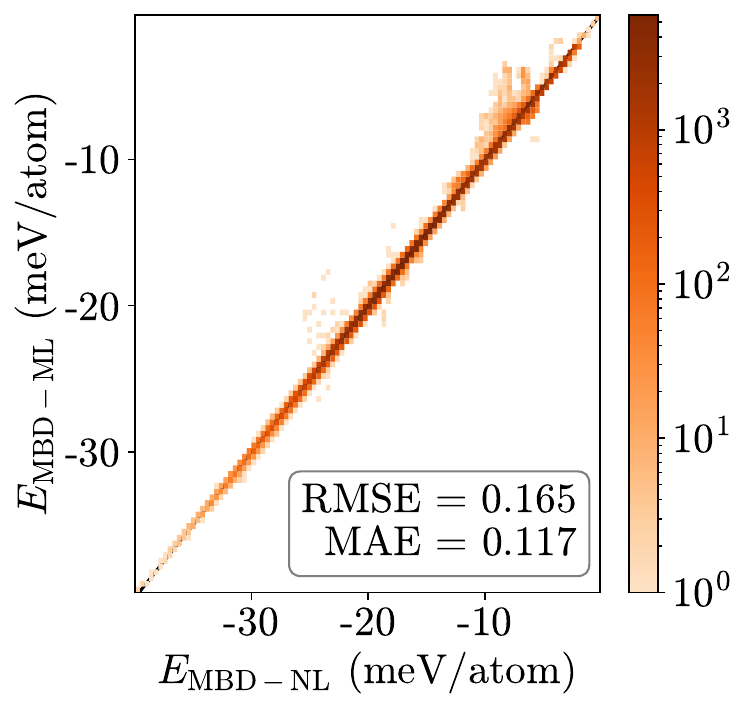}
\end{subfigure}%
\begin{subfigure}{0.48\linewidth}
\caption{}
\includegraphics[width=\linewidth]{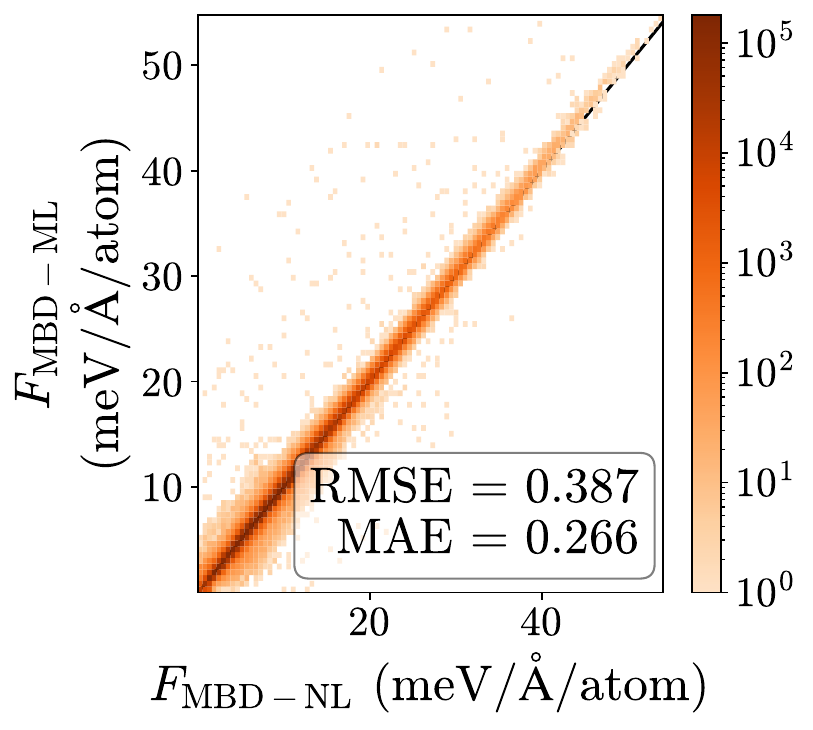}
\end{subfigure}
\caption{Performance of PBE0+MBD-ML in predicting the $C_6$
and $\alpha_0$ ratios and the MBD contribution to the total energy and atomic forces of the DES370k test set excluding alkali and alkaline earth elements}
\label{fig:c6-a0-e-f-performance-des370k-no-alkali}
\end{figure}

While the main analysis of the MBD-ML performance on the OMol25 and DES370k subsets
excluded alkali and alkaline earth elements, here we additionally show the accuracy 
of the model if those elements are included relative to the \emph{ab initio} MBD-NL reference.
Figure~\ref{fig:c6-a0-e-f-performance-omol25} and~\ref{fig:c6-a0-e-f-performance-omol25-no-alkali} illustrate the performance of the MBD-ML model on the OMol25 subset with and without alkali and alkali earth elements. Figure~\ref{fig:c6-a0-e-f-performance-des370k} and~\ref{fig:c6-a0-e-f-performance-des370k-no-alkali} do so analogously for the DES370k test set.

\clearpage

\section{Sensitivity analysis of MBD-ML}

The present study demonstrates excellent accuracy of the MBD-ML model for nearly all target quantities, including relative energies, atomic forces, and stress tensor components. However, as noted in the main text and evident from Fig.~4, the total MBD energy predicted by MBD-ML can deviate from the \emph{ab initio} reference by more than $100,\mathrm{meV}$ for large systems containing over one hundred atoms. As shown in Table~III, this discrepancy becomes negligible when computing physically measurable energy differences. Nevertheless, understanding the origin of this residual disagreement is instructive. In principle, exact prediction of the $\alpha_0^{\mathrm r}$ and $C_6^{\mathrm r}$ ratios would eliminate this error entirely.

To assess the sensitivity of the MBD energy to these inputs, we performed a sensitivity analysis for polymorph I of coronene. The MBD-ML predicted vdW energy for coronene I differs from the MBD-NL reference by $-61,\mathrm{meV}$ ($-0.85,\mathrm{meV/atom}$), consistent with the MAE of the MBD-ML model on the OMC25 subset (Table~I).

\begin{figure}
    \centering
    \includegraphics[width=\linewidth]{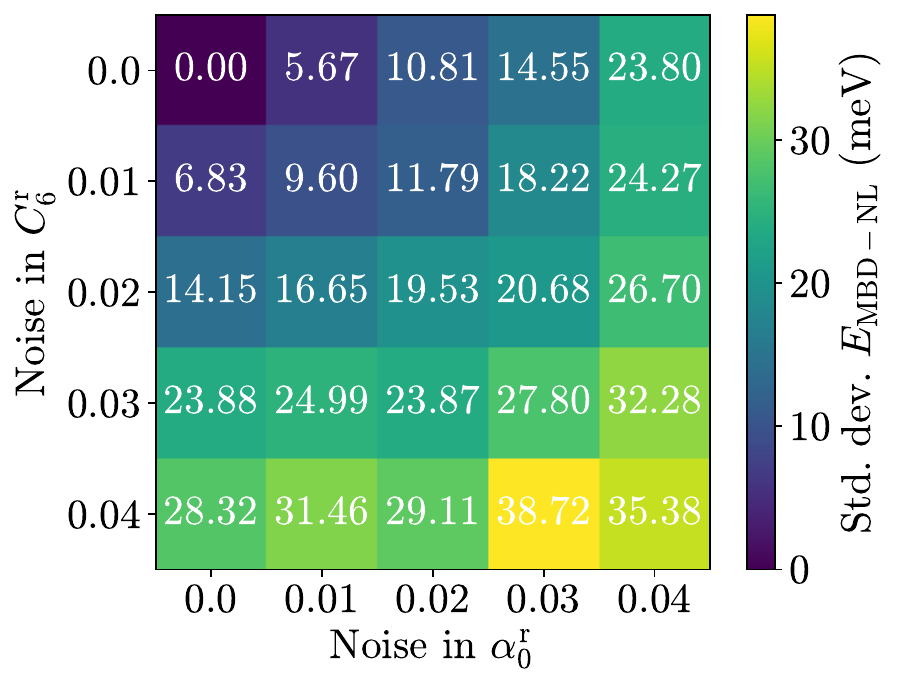}
    \caption{Sensitivity of the MBD-NL energy to the
    introduction of uniform random noise to
    $a_0^{\mathrm{r}}$ and $C_6^{\mathrm{r}}$}
    \label{fig:rdm_noise_sens_analysis}
\end{figure}

We first examined the impact of random, non-systematic noise in $\alpha_0^{\mathrm r}$ and $C_6^{\mathrm r}$ on the total MBD energy. Starting from the PBE0+MBD-NL reference ratios of coronene I, uniform zero-mean noise with a variable range between $0.0$ and $0.04$ was applied independently to both ratios. For each noise level combination, the MBD-NL energy was recomputed 100 times, and the resulting standard deviation was recorded (Fig.~\ref{fig:rdm_noise_sens_analysis}). The chosen range reflects the average prediction errors for these ratios on the OMC25 subset (Table~I).

Perturbations within this range produce standard deviations in the total MBD energy of up to $\pm 38,\mathrm{meV}$, substantially smaller than the $-61,\mathrm{meV}$ deviation observed for MBD-ML on coronene I. Based on Fig.~\ref{fig:rdm_noise_sens_analysis} and the reported MAEs, the non-systematic contribution to the MBD energy error is estimated to be about $20,\mathrm{meV}$. The analysis also indicates that the MBD energy is slightly more sensitive to noise in $C_6^{\mathrm r}$ than in $\alpha_0^{\mathrm r}$, as evidenced by consistently larger deviations in the lower-left region of the heat map, although the difference amounts to only a few meV.

\begin{figure}
    \centering
    \includegraphics[width=\linewidth]{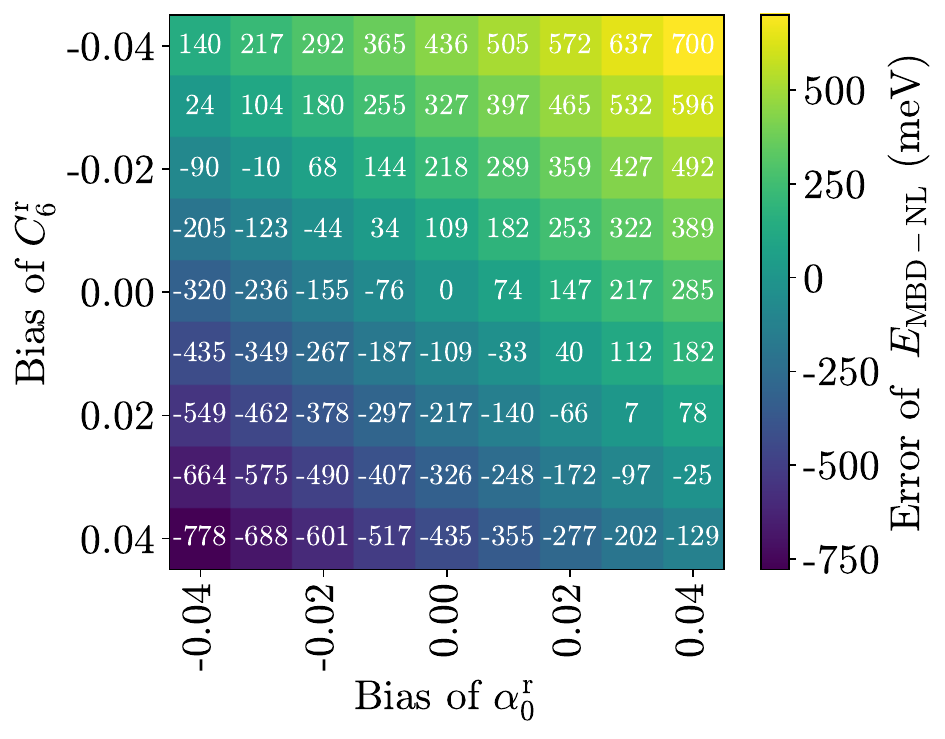}
    \caption{Sensitivity of the MBD-NL energy to the
    introduction of a systematic bias to
    $a_0^{\mathrm{r}}$ and $C_6^{\mathrm{r}}$}
    \label{fig:sys_bias_sens_analysis}
\end{figure}

In addition to random errors, systematic bias must be considered. For the 200 molecular crystals in the OMC25 dataset, the mean residuals (serving as a proxy for bias) in Fig.~4 are $-0.02$ for $\alpha_0^{\mathrm r}$ and $-0.01$ for $C_6^{\mathrm r}$. To quantify the effect of such bias, the ratios for coronene I were independently shifted by values between $-0.04$ and $0.04$ (Fig.~\ref{fig:sys_bias_sens_analysis}). The resulting energy errors reveal a much stronger sensitivity to systematic shifts, with deviations reaching up to $700,\mathrm{meV}$ and $-778,\mathrm{meV}$. Again, shifts in $C_6^{\mathrm r}$ exert a larger influence on the vdW energy than comparable shifts in $\alpha_0^{\mathrm r}$.

Interestingly, energy errors are reduced when shifts in both ratios are of similar magnitude. Points along the diagonal of Fig.~\ref{fig:sys_bias_sens_analysis}, where biases are balanced, exhibit smaller deviations. The observed biases of $-0.02$ and $-0.01$ ((Table~I) correspond to an estimated MBD energy error of about $-44,\mathrm{meV}$. Combining this systematic contribution with the estimated non-systematic uncertainty yields a total error estimate of $-44 \pm 20,\mathrm{meV}$, consistent with the observed $-61,\mathrm{meV}$ deviation for coronene I.

Beyond rationalizing residual errors, this analysis provides guidance on the accuracy requirements for future models. Sensitivity estimates of this type can be used to determine the level of precision in predicted vdW ratios necessary to achieve a desired accuracy in the resulting MBD energies.

\clearpage
\section{Details on force deviation distribution for MBD-ML, D3 and D4}
\begin{figure}[h]
\centering
    \includegraphics[width=0.9\linewidth]{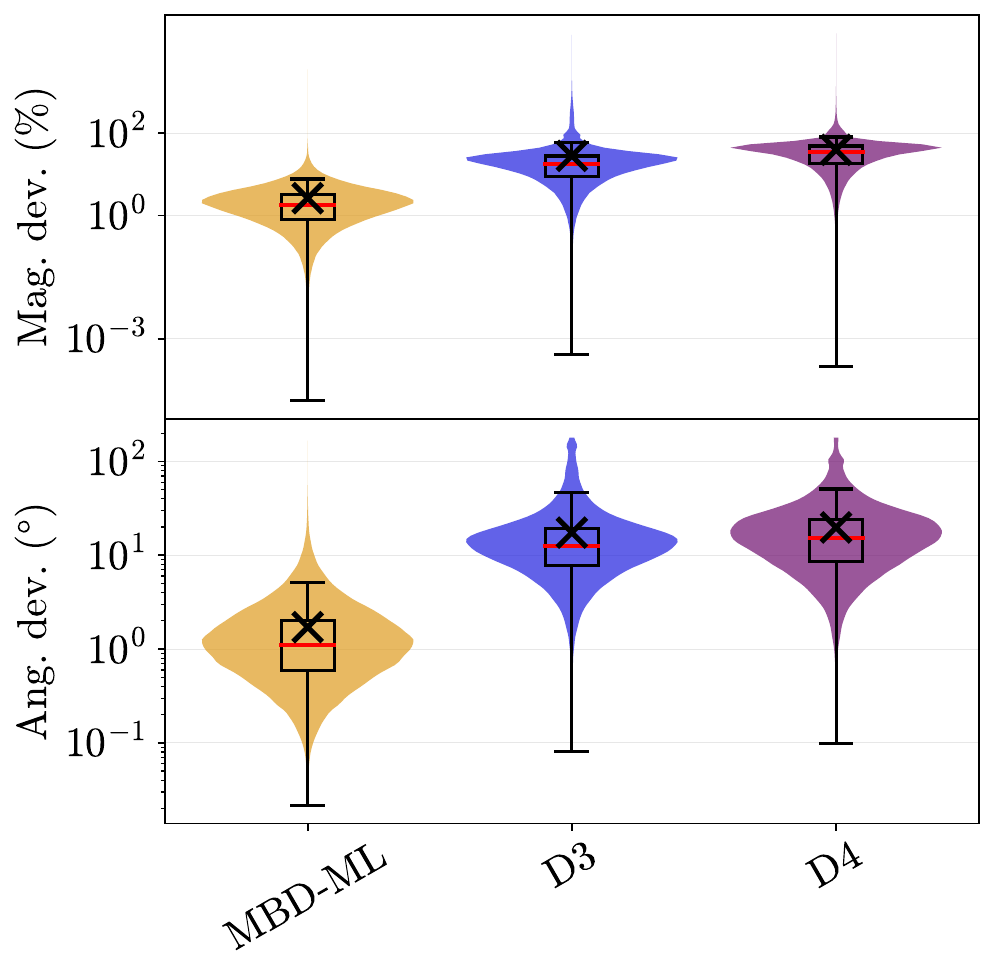}
    \caption{Distributions of vdW force deviations from the MBD-ML, D3 and D4 method compared to MBD-NL. The distributions are represented by violin plots. Black crosses correspond to the mean value, the red line to the median. The vertical edges of the boxes are the quartiles, i.e the 25-th and 75-th percentile of the deviation distribution. The whiskers correspond to the minimum and the $95\%$ deviation}
    \label{fig:violin-plots}
\end{figure}

Here we provide the quantitative statistical data on the distribution of the force magnitudes and
angles of the MBD-ML, the D3 and the D4 method relative to the \emph{ab initio} MBD-NL method.
D3 and D4 vdW force contributions were computed using the \texttt{dftd3} (\url{https://github.com/dftd3/simple-dftd3}) and \texttt{dftd4} (\url{https://github.com/dftd4/dftd4}) packages. Both corrections were computed using the default PBE0 parameters, included Becke-Johnson (BJ) damping and 
a Axilrod-Teller-Muto three-body term and were compared to PBE0-based MBD-NL forces with a damping parameter $\beta=0.83$. The same $\beta$ was used for the MBD-ML calculations.
The distributions are illustrated via violin plots in Figure~\ref{fig:violin-plots}, exact 
statistical values are given in Table~\ref{tab:violin-plot-stats}.

\begin{table}
\caption{Mean, median, and 95\textsuperscript{th} percentile deviations for relative magnitude and angular force predictions of MBD-ML, D3 and D4 compared to MBD-NL on the OMol25 test set}
\centering
\setlength{\tabcolsep}{12pt}
\begin{tabular}{l ccc}
\toprule
& \multicolumn{1}{c}{MBD-ML} & \multicolumn{1}{c}{D3} & \multicolumn{1}{c}{D4}\\
\midrule
\multicolumn{3}{l}{\textit{Magnitude Deviation (\%)}}\\[2pt]
\quad Mean   & 2.6 & 28.2& 39.5\\
\quad Median & 1.8 & 18.2& 35.4\\
\quad 95\textsuperscript{th} Percentile & 7.7& 60.7 &80.8\\
\midrule
\multicolumn{3}{l}{\textit{Angular Deviation (°)}}\\[2pt]
\quad Mean   & 1.7 & 17.4 & 19.8\\
\quad Median & 1.1 & 12.6 & 15.2\\
\quad 95\textsuperscript{th} Percentile & 5.1 & 46.6 & 50.8\\
\bottomrule
\end{tabular}
\label{tab:violin-plot-stats}
\end{table}

\section{Geometry optimization of molecular crystals}

  \begin{table}[h]
 \centering
 \begin{tabular}{llc}
 \toprule
 Molecule & Polymorphs & interaction type \\
 \midrule\midrule
 aspirin & \makecell[l]{I (ACSALA07)\\ II (ACSALA17)}& mixed\\
 \midrule
 maleic acid & \makecell[l]{I (MALIAC12)\\ II (MALIAC13)}& hb\\
 \midrule
 2-amino-5-nitropyrimidine & \makecell[l]{I (PUPBAD01)\\ II (PUPBAD02)\\ III (PUPBAD)} & hb\\
 \midrule
 maleic hydrazide & \makecell[l]{ I (MALEHY10)\\ II (MALEHY01)\\ III (MALEHY12)}& hb\\
 \midrule
 pentacene & \makecell[l]{I (PENCEN)\\ II (PENCEN04)}& vdW\\
 \midrule
 coronene & \makecell[l]{I (CORONE04$^{\mathrm{a}}$)\\ II (N/A$^{\mathrm{a}}$)}&vdW\\
 \midrule
 o-diiodobenzene&\makecell[l]{I (ZZZPRO06)\\ II (ZZZPRO07)}&vdW\\
 \bottomrule
 \bottomrule
 \end{tabular}
 \smallskip\raggedright
 $^{\mathrm{a}}$ Structures directly taken from Ref~\citenum{potticary2016unforeseen}.
 \caption{Molecular crystal polymorphs from which geometry optimizations were performed. Along with the roman numeral denoting the polymorph the CSD reference code and the type of interaction predominantly found in the system is provided: mixed, van-der-Waals (vdW) or hydrogen-bonded (hb)}
 \label{tab:geo-op-molcrystals}
 \end{table}

For the structure relaxation of the molecular crystals
the \texttt{Aims} calculator of the Atomic Simulation Environment (ASE) was employed. For geometry optimizations that involved the MBD-ML model, 
a new ASE calculator was implemented
and combined with the Aims calculator as a \texttt{SumCalculator}, so that
the final calculator object yielded total energies, atomic forces and stress tensors which were the sum of the pure FHI-aims output
and the MBD-ML contributions to these properties.
An example script to perform a PBE0+MBD-ML geometry relaxation can be found in the libmbd repository under \url{https://github.com/em819/libmbd/tree/master/examples/mbd-ml-ase/aspirin-structure-relaxation}.
All geometry optimizations of molecular crystals were performed with the
ASE implementation of the BFGS algorithm and involved two steps:
Starting from the CSD reference structure, in the first step only the 
atomic positions were relaxed with a force
threshold \texttt{fmax} of $0.005\,\text{eV/\AA}$. The resulting structure was
relaxed again, including both the atomic positions and the unit cell
using the \texttt{FrechetCellFilter} without any constraints to a \texttt{fmax} of $0.005\,\text{eV/\AA}$.
In this second step, the initial guess for the Hessian in the ASE BFGS algorithm,
\texttt{alpha} was chosen as $150.0$ instead of the default $70.0$
to counteract overshooting due to the notoriously shallow and rugged potential energy surface of molecular crystals, leading
to more stable convergence. 
While the geometry optimizations were performed with the PBE DFA with MBD-NL,
MBD-ML and without and vdW method, the final optimized structure was recomputed
with the PBE0 functional instead. For PBE-based calculations that were combined with a the MBD method (MBD-NL or MBD-ML), a damping parameter $\beta$ of 0.81 was used, while for the final PBE0 recalculations, a value of 0.83 was set.

\begin{table*}
\caption{Structural comparison of polymorphs relaxed with PBE and PBE+MBD-ML relative to the PBE+MBD-NL-relaxed reference structures. RMSD and maximum atomic displacement (Max.\ Disp.) quantify the deviation of atomic positions, while $\Delta V$ gives the relative change in unit cell volume. Cases where the PBE geometry optimization did not converge are marked with ``--''. ANP refers to 2-amino-5-nitropyrimidine}
\centering
\setlength{\tabcolsep}{4pt}
\begin{tabular}{l l cccc cc}
\toprule
 & & \multicolumn{2}{c}{RMSD (\AA)} & \multicolumn{2}{c}{Max.\ Disp.\ (\AA)} & \multicolumn{2}{c}{$\Delta V$ (\%)} \\
\cmidrule(lr){3-4} \cmidrule(lr){5-6} \cmidrule(lr){7-8}
Molecule & Polymorph & PBE & PBE+MBD-ML & PBE & PBE+MBD-ML & PBE & PBE+MBD-ML \\
\midrule
\multicolumn{8}{l}{\textit{Mixed Interactions}}\\[2pt]
\quad Aspirin & I & 0.177 & 0.002 & 0.248 & 0.003 & 26.00 & 0.09 \\
\quad & II & -- & 0.007 & -- & 0.018 & -- & 0.22 \\
\midrule
\multicolumn{8}{l}{\textit{Hydrogen Bonded}}\\[2pt]
\quad Maleic acid & I & 0.035 & 0.002 & 0.060 & 0.003 & 24.16 & 0.29 \\
\quad & II & 0.040 & 0.001 & 0.077 & 0.002 & 27.19 & 0.19 \\
\quad ANP & I & 0.222 & 0.004 & 0.385 & 0.006 & 28.32 & 0.22 \\
\quad & II & 0.401 & 0.002 & 1.046 & 0.004 & 37.47 & -0.18 \\
\quad & III & 0.062 & 0.001 & 0.149 & 0.001 & 13.22 & 0.05 \\
\quad Maleic hydrazide & I & 0.055 & 0.001 & 0.078 & 0.002 & 28.13 & 0.05 \\
\quad & II & 0.170 & 0.003 & 0.196 & 0.007 & 30.76 & -0.11 \\
\quad & III & 0.042 & 0.001 & 0.068 & 0.002 & 23.23 & 0.00 \\
\midrule
\multicolumn{8}{l}{\textit{van der Waals}}\\[2pt]
\quad Pentacene & I & 0.176 & 0.004 & 0.293 & 0.008 & 36.18 & -0.46 \\
\quad & II & -- & 0.007 & -- & 0.014 & -- & -0.55 \\
\quad Coronene & I & -- & 0.004 & -- & 0.007 & -- & -0.26 \\
\quad & II & 0.277 & 0.001 & 0.499 & 0.002 & 37.42 & -0.18 \\
\quad \textit{o}-Diiodobenzene & I & 0.119 & 0.005 & 0.179 & 0.008 & 29.86 & -0.50 \\
\quad & II & 0.085 & 0.038 & 0.119 & 0.056 & 26.65 & 0.12 \\
\bottomrule
\end{tabular}
\label{tab:si-rmsd-vol-ratio}
\end{table*}

\pagebreak 

\bibliography{references}